\documentclass[12pt]{article}
\pdfoutput=1
\usepackage{hyperref}
\usepackage{amsmath,amssymb,amsthm,amscd}
\usepackage{mathtools}
\usepackage{bm}
\usepackage{graphicx}

\newcommand{\dd}{\mathrm{d}}
\newcommand{\Z}{\mathbb{Z}}
\newcommand{\R}{\mathbb{R}}

\newcommand{\Tr}{\textrm{Tr}}

\newcommand{\be}{\begin{equation}}
\newcommand{\ee}{\end{equation}}
\newcommand{\ba}{\begin{aligned}}
\newcommand{\ea}{\end{aligned}}

\numberwithin{equation}{section}

\begin{document}
\begin{titlepage}

\rightline{USTC-ICTS/PCFT-26-65}

\vskip 3 cm

\centerline{\Large 
\bf  
AI for BMN Strings   }

\vskip 0.5 cm

\renewcommand{\thefootnote}{\fnsymbol{footnote}}
\vskip 30pt \centerline{ {\large \rm 
Min-xin Huang\footnote{minxin@ustc.edu.cn}  
} } \vskip .5cm  \vskip 20pt 

\begin{center}
{Interdisciplinary Center for Theoretical Study,  \\ \vskip 0.1cm  University of Science and Technology of China,  Hefei, Anhui 230026, China} 
 \\ \vskip 0.3 cm
{Peng Huanwu Center for Fundamental Theory,  \\ \vskip 0.1cm  Hefei, Anhui 230026, China} 
\end{center}

\setcounter{footnote}{0}
\renewcommand{\thefootnote}{\arabic{footnote}}
\vskip 40pt
\begin{abstract}

With the assistance of AI (Artificial Intelligence) tools, we revisit some mathematically well-formulated problems from our previous studies of BMN (Berenstein-Maldacena-Nastase) strings on type IIB string theory pp-wave background. These mathematical problems have important physical implications, relating to the computations of higher-genus critical superstring amplitudes, possible estimate of the cosmological constant. We make significant progress on some problems, while others remain unresolved.

\end{abstract}

\end{titlepage}
\vfill \eject


\newpage

\baselineskip=16pt

\tableofcontents

\section{Introduction}

Recently the capability of AI models for solving mathematical problems has increased significantly. In particular, a historic achievement is the solution of a Millennium prize problem on Navier-Stokes equations by an advanced AI model, based on some earlier human works. At present, the AI tools are particularly suited for studying mathematically well-formulated problems. It would be interesting to apply the tools to such problems with important  physical implications. 

The AdS/CFT correspondence provides a non-perturbative definition of quantum gravity on AdS background in terms of a conformal field theory (CFT) on the boundary \cite{Maldacena:1997re}. In this paper we focus on type IIB string theory whose dual is the 4d $\mathcal{N}=4$ $SU(N)$ super-Yang-Mills theory. In the Penrose limit, the AdS background becomes the pp-wave background \cite{Blau:2001ne} and the closed string states are described by the BMN operators in the dual CFT \cite{Berenstein:2002jq}. We denote the BMN operators by $O^J_{n_1, n_2, \cdots, n_k}$, which is constructed by inserting $k$ distinct string modes into the vacuum operator $\Tr(Z^J)$ with phases, where $Z$ is one of the complex scalar of the $\mathcal{N}=4$  super-Yang-Mills theory. Due to the closed string level matching condition, we have the sum of the integer string modes $\sum_i n_i =0$ and the excited stringy states have at least two string modes. In the BMN double scaling limit, we send $N, J\rightarrow \infty$ and keep the genus counting parameter $g=\frac{J^2}{N}$ fixed. 

We further focus on the infinitely curved pp-wave background which is dual to free gauge theory, where the string spectrum is completely degenerate.  In this case, the holographic correspondence is manifested through explicit relations and patterns in the free BMN correlators.  In our previous works \cite{Huang:2002wf, Huang:2002yt, Huang:2010ne, Huang:2019lso, Huang:2019uue, Du:2021dml, Du:2021spv, Huang:2025dnq}, we proposed and studied such mathematically well-defined relations and their physical implications. In this paper, with the assistance of public tools from OpenAI ChatGPT, we will make progress on some previous conjectures.
 
There are some important physical implications. A long standing problem in string theory is the computations of higher-genus critical superstring amplitudes. It is generally accepted that the physical definition is UV finite with no pathology or ambiguity. However, there are still some subtleties in the higher genus moduli space so a complete rigorous mathematical definition is not available and practical calculations are also quite difficult.  A technical difficulty is the picture changing operation introduced in the original formulation \cite{Friedan:1985ge}. For some heroic efforts on this issue see e.g. \cite{Sen:2015hia}. For comparison, the simpler topological string theory has a rigorous mathematical definition as Gromov-Witten theory and the amplitudes on compact Calabi-Yau spaces can be computed by physical methods to very high genus \cite{Candelas:1990rm, Bershadsky:1993cx, Huang:2006hq, Alexandrov:2023zjb}.  The situation is also easier in low dimensional non-critical strings, see e.g. \cite{Balthazar:2019rnh}. If our conjectures for the ``pp-wave holographic dictionary" are correct, we have provided the critical superstring amplitudes at any higher genus, and even exact results at any finite string coupling due to the convergence of the genus expansion, in the extreme pp-wave background with infinite spacetime curvature and Ramond-Ramond flux. 

Another physical application concerns cosmology. Recently, we made a concrete testable proposal to use BMN strings as a probe of the entropy of our universe, which could provide a possible estimate of the cosmological constant \cite{Huang:2025dnq}. Unfortunately, in Section \ref{subsecunbounded}, we prove with AI assistance that the entropy of BMN strings is unbounded in the strong coupling limit, rendering this proposal nonviable.

\section{Proof of the factorization formulas}

The factorization formulas were first proposed in \cite{Huang:2002yt} and later completely formulated in \cite{Huang:2010ne}. A small discrepancy in the $2\rightarrow 2$ process was also later resolved \cite{Huang:2019lso} as missing diagrams in the calculations. The free planar three-point functions of BMN operators are identified with the Green-Schwarz light-cone string-field-theory cubic vertices in the infinite curvature limit \cite{Spradlin:2002ar, Huang:2002wf}. Some early computations of non-planar BMN correlators can be found in \cite{Constable:2002hw}. A string diagram is constructed by pasting together the cubic vertices without propagators. On the other hand, a field-theory diagram calculates the free BMN correlator on the gauge theory side by Wick contractions. Denote the string- and field-theory diagram contributions by $S_i$ and $F_j$. Let $m_{ij}$ count the number of times string diagram $i$ appears when the short process of field-theory diagram $j$ is expanded into long processes. The factorization formulas are 
\be 
S_i =\sum_j m_{i j} F_j . 
\ee

The factorization formulas are checked in various examples in \cite{Huang:2002yt, Huang:2010ne, Huang:2019lso, Du:2021dml}. We give below a general derivation,  for simplicity  focusing on the case where both the incoming and outgoing states are single-trace BMN operators. AI models can also prove the general cases with more complicated arguments.

In the physical setting, the four remaining real scalars of the $\mathcal{N}=4$ super-Yang-Mills theory provide string modes in the four transverse dimensions of the Ramond-Ramond flux. Including covariant derivatives, we have at most eight distinct string modes. However, the mathematical structures are stronger that the same combinatorial structure applies to any number of string modes.  

\subsection{Genus one}

For illustration we first consider the simple case of genus one. In this case there are only one string and one field theory diagram. The field theory short process $(1234)\rightarrow (2143)$  can be expanded into two long processes 
\be \ba 
&& (1234)\rightarrow (12)(34) \rightarrow  (2143), \\
&& (1234)\rightarrow (41)(23) \rightarrow  (1432), 
\ea \ee
where the permutation is equivalent by cyclic rotations. So the multiplicity is $2$ and the factorization formula is just $S=2F$. The relation was checked for the cases of two \cite{Huang:2002yt}, three \cite{Huang:2010ne}, four \cite{Du:2021dml} string modes. We now give a proof for any number $k$ of string modes.

 Let $ {\bm m}=(m_1,\ldots,m_k),  {\bm n}=(n_1,\ldots,n_k)\in\Z^k$ be two sets of string modes, satisfying the level matching condition 
\[
 \sum_{i=1}^k m_i=0,\qquad \sum_{i=1}^k n_i=0.
\]
Then the factorization formula is 
\be
 \sum_{\text{two-string intermediate states}}
 \big\langle \overline{O}_{\bm m}\,O_{\bm p} O_{\bm q}
 \big\rangle_{\mathrm{planar}}\,
 \big\langle \overline{O}_{\bm p} \overline{O}_{\bm q} \,
 O_{\bm n}\big\rangle_{\mathrm{planar}} =  2 \big\langle  \overline{O}_{\bm m} O_{\bm n}\big\rangle_{\mathrm{torus}} .
 \label{tag1}
\ee
The intermediate sum includes the choice of size and a subset of $k$ string modes for the daughter strings, and we also need to sum over all intermediate mode numbers ${\bm p}$ and ${\bm q}$ satisfying the level matching condition. 

Choose a subset \(A\subseteq\{1,\ldots,k\}\), and put
\[
 B=A^{c},\qquad a=|A|,\qquad b=|B|,\qquad a+b=k.
\]
Suppose the parent string size is $J$, the first daughter string size is \(x J \), and the second string size is \((1-x)J \), where \(0<x<1\).  Let
\(\bm p=(p_i)_{i\in A}\) and \(\bm q=(q_j)_{j\in B}\) be the mode labels on
the two daughter strings, with
\[
 \sum_{i\in A}p_i=0,\qquad \sum_{j\in B}q_j=0.
\]
The planar three-string vertex can be written as
\begin{align}
 V_{\bm m}^{A;\bm p,\bm q}(x)
 ={}&
 \frac{g}{\sqrt J}\,
 x^{(1-a)/2}(1-x)^{(1-b)/2}
 \nonumber\\
 &\times
 \prod_{i\in A}\int_0^x \dd y_i\,
 \exp\!\left(-2\pi i m_i y_i+\frac{2\pi i p_i}{x}y_i\right)
 \nonumber\\
 &\times
 \prod_{j\in B}\int_x^1 \dd y_j\,
 \exp\!\left(-2\pi i m_j y_j
 +\frac{2\pi i q_j}{1-x}(y_j-x)\right).
 \label{tag2}
\end{align}
The string diagram contribution obtained by gluing two such vertices is
\begin{align}
 S
 ={}&
 \frac{J}{2}\int_0^1\dd x
 \sum_{A\subseteq\{1,\ldots,k\}}
 \sum_{\substack{\bm p:\,\sum_{i\in A}p_i=0\\
                  \bm q:\,\sum_{j\in B}q_j=0}}
 V_{\bm m}^{A;\bm p,\bm q}(x)\,
 \overline{V_{\bm n}^{A;\bm p,\bm q}(x)} .
 \label{tag3}
\end{align}
The factor \(1/2\) removes the double counting obtained by interchanging the
two daughter strings:
\[
 (A,x,\bm p,\bm q)\longleftrightarrow
 (B,1-x,\bm q,\bm p).
\]

To sum the internal modes, we define the periodic delta distribution for a circle of circumference \(\ell\), 
\[
 \delta_\ell(z)=\sum_{r\in\Z}\delta(z-r\ell).
\]
The constrained Fourier-completeness identity is
\begin{align}
 \sum_{\substack{\bm p\in\Z^r\\ \sum_{a } p_a=0}}
 \exp\!\left[
 \frac{2\pi i}{\ell}\sum_{a=1}^r p_a(u_a-v_a)
 \right]
 =
 \ell^{\,r-1}\int_0^\ell \dd\sigma\,
 \prod_{a=1}^r
 \delta_\ell(u_a-v_a-\sigma).
 \label{tag4}
\end{align}
This can be derived by using Poisson summation formula for $ \delta_\ell $ and the integral imposes the level-matching constraint $\sum_{a } p_a=0$.

Apply (\ref{tag4}) to the modes on the first daughter string, whose circumference is
\(x\), and to the modes on the second daughter string, whose circumference is
\(1-x\).  Denote the corresponding translation parameters by
\(t\in[0,x]\) and \(u\in[0,1-x]\).  The powers of \(x\) and \(1-x\) in
(\ref{tag2}) cancel the powers produced by (\ref{tag4}), and the sum over intermediate modes
therefore gives
\begin{align}
 S
 ={}&
 \frac{g^2}{2}
 \int_0^1\dd x\int_0^x\dd t\int_0^{1-x}\dd u
 \sum_{A\subseteq\{1,\ldots,k\}}
 \int_{D_A}\prod_{i=1}^k\dd y_i
 \nonumber\\
 &\qquad\qquad\times
 \exp\!\left[
 2\pi i\sum_{i=1}^k
 \bigl(n_iF_{x,t,u}(y_i)-m_i y_i\bigr)
 \right],
 \label{tag5}
\end{align}
where
\be
 D_A=
 \left(\prod_{i\in A}[0,x]\right)
 \times
 \left(\prod_{j\in B}[x,1]\right),
 \label{tag6}
\ee
and \(F_{x,t,u}\) is the piecewise translation
\be
 F_{x,t,u}(y)=
 \begin{cases}
  (y+t)\bmod x, & 0\le y<x,\\[2mm]
  x+\bigl((y-x+u)\bmod(1-x)\bigr), & x\le y<1.
 \end{cases}
 \label{tag7}
\ee

The domains \(D_A\) form a disjoint partition of the unit cube
\([0,1]^k\), up to measure-zero boundaries:
\[
 [0,1]^k=\bigsqcup_{A\subseteq\{1,\ldots,k\}}D_A.
\]
Consequently, (\ref{tag5}) becomes
\begin{align}
 S
 ={}&
 \frac{g^2}{2}
 \int_0^1\dd x\int_0^x\dd t\int_0^{1-x}\dd u
 \int_{[0,1]^k}\prod_{i=1}^k\dd y_i
 \nonumber\\
 &\qquad\qquad\times
 \exp\!\left[
 2\pi i\sum_{i=1}^k
 \bigl(n_iF_{x,t,u}(y_i)-m_i y_i\bigr)
 \right].
 \label{tag8}
\end{align}

We now change the integration variables to obtain a simplex integral of the torus two point function.  Introduce four nonnegative variables by
\be
 x_1=x-t,\qquad
 x_2=t,\qquad
 x_3=1-x-u,\qquad
 x_4=u.
 \label{tag9}
\ee
Then
\[
 x_r\ge0,\qquad
 x_1+x_2+x_3+x_4=1,
 \qquad
 x=x_1+x_2,\quad t=x_2,\quad u=x_4.
\]
The map has unit Jacobian, and (\ref{tag8}) can therefore be written as
\begin{align}
 S
 ={}&
 \frac{g^2}{2}
 \int_{\Delta_3}\dd^3\bm x
 \int_{[0,1]^k}\prod_{i=1}^k\dd y_i\,
 \exp\!\left[
 2\pi i\sum_{i=1}^k
 \bigl(n_iF_{\bm x}(y_i)-m_i y_i\bigr)
 \right],
 \label{tag10}
\end{align}
where
\be
 \Delta_3=
 \left\{\bm x\in\R_{\ge0}^4:
 \sum_{r=1}^4x_r=1\right\},
 \qquad
 \dd^3\bm x:=\dd x_1\dd x_2\dd x_3,
 \quad x_4=1-x_1-x_2-x_3,
 \label{tag11}
\ee
and \(F_{\bm x}\) denotes (\ref{tag7}) after the substitution (\ref{tag9}).

It remains to put \(F_{\bm x}\) into the standard four-segment form.  On the
four source intervals
\[
 [0,x_1],\quad [x_1,x_1+x_2],\quad
 [x_1+x_2,1-x_4],\quad [1-x_4,1],
\]
the differences \(F_{\bm x}(y)-y\) are respectively
\be
 t,\qquad t-x,\qquad u,\qquad u-(1-x).
 \label{tag12}
\ee
Subtracting the common translation \(t\) from \(F_{\bm x}\) changes the
exponent in (\ref{tag10}) by
\[
 -2\pi i\,t\sum_{i=1}^k n_i=0.
\]
On the second interval one may additionally add \(1\) to the translated
value of \(F_{\bm x}\), because every \(n_i\) is an integer.  Thus the phase
is unchanged if \(F_{\bm x}\) is replaced by the map \(P_{\bm x}\) whose
displacement is
\begin{align}
 P_{\bm x}(y)-y
 =
 \begin{cases}
  0, & 0\le y<x_1,\\
  x_3+x_4, & x_1\le y<x_1+x_2,\\
  x_4-x_2, & x_1+x_2\le y<1-x_4,\\
  -(x_2+x_3), & 1-x_4\le y<1.
 \end{cases}
 \label{tag13}
\end{align}
For example, before the integer shift the second displacement is
\(-x=-(x_1+x_2)\); adding \(1\) gives
\(1-x=x_3+x_4\).  The other three displacements follow directly from
(\ref{tag12}) and (\ref{tag9}). Hence
\begin{align}
 S
 ={}&
 \frac{g^2}{2}
 \int_{\Delta_3}\dd^3\bm x
 \prod_{i=1}^k
 \left[
 \int_0^1\dd y\,
 \exp\!\left(
 2\pi i\bigl[n_iP_{\bm x}(y)-m_i y\bigr]
 \right)
 \right].
 \label{tag14}
\end{align}
The product notation is valid because the \(y_i\)-integrals are independent
once the common map \(P_{\bm x}\) has been fixed. This matches exactly twice the torus two-point function, see e.g. \cite{Du:2021dml}. So we have 
$S = 2 \big\langle  \overline{O}_{\bm m} O_{\bm n}\big\rangle_{\mathrm{torus}}$,  which is exactly the factorization statement (\ref{tag1}).

\subsection{General arguments}

We consider a fixed genus $h\geq1$, with a single-trace operator at
each endpoint; the planar two-point function is the identity.  The
argument has two parts.  First we count the splitting-and-joining
histories that produce a given field-theory diagram.  Then we show
that the sewing variables for each history can be replaced by the
lengths of its $4h$ segments, giving the same simplex integral as on
the field-theory side.  We keep the number $k$ of distinct impurities
and the external modes fixed in the strict BMN limit, and begin with
states of definite impurity positions at finite $J$.

\medskip
\noindent\emph{The cuts of a leading contraction.}
Temporarily label the $Q=J+k$ fields on the incoming trace.  Write
$\alpha(a)$ for the label following $a$.  After transporting the
labels through a Wick contraction, write $\beta(a)$ for the outgoing
successor, with a consistent orientation.  Following the double-line
indices gives one color loop for each cycle of
$\rho=\alpha^{-1}\beta$.  Thus Euler's formula is
\begin{equation}
 Q-c(\alpha^{-1}\beta)=2h,
 \label{eq:factorization-euler-defect}
\end{equation}
where $c$ includes fixed points.  If the nontrivial cycles of $\rho$
have lengths $r_a$, the number $d$ of changed successor links obeys
\begin{equation}
 d=\sum_a r_a\leq2\sum_a(r_a-1)=4h.
 \label{eq:factorization-support-bound}
\end{equation}
Each distinct cut supplies a freely summed vacuum position.  Terms
with $d<4h$ lose at least one power of $J$ and are subleading; fixed
$k$ and the normalized impurity sums do not change this counting.
The leading contractions therefore have $4h$ cuts.  Equality in
\eqref{eq:factorization-support-bound} means that $\rho$ consists of
$2h$ disjoint transpositions, which pair these cuts.  Repeated cuts
and cuts at impurities are suppressed in the same way.  Here we use
the $U(N)$ propagator; the $SU(N)$ trace subtractions are subleading
in the BMN limit.

A cubic vertex exchanges two successor links:
\begin{equation}
 \gamma\longmapsto\gamma(a\,b).
 \label{eq:factorization-successor-swap}
\end{equation}
It cuts after $a$ and $b$ and exchanges their successors.  This splits
one cyclic word when the cuts are on the same string, and joins two
words otherwise.  It is the cut-and-join description of the planar
three-point vertex \cite{Brown:2010pb}.  A genus-$h$ string diagram
between two single strings has $2h$ cubic vertices, so a leading
history uses each of the $2h$ paired cuts once.

\medskip
\noindent\emph{How to determine $m_{ij}$.}
Fix a field-theory short process $j$, labeling the incoming segments
$1,\ldots,4h$ in cyclic order.  Its outgoing order fixes the paired
reconnections just described.  To find its multiplicity in a string
diagram $i$, assign these pairs to the vertices of $i$ and follow the
resulting cyclic words along its edges.  Keep an assignment precisely
when every vertex acts on the strings specified by $i$ and the final
word is the outgoing word of $j$.  Checking the numbers of strings
alone is insufficient: the connections between vertices must also
agree with $i$.  All segment labels are retained at intermediate
steps, even if neighboring segments could temporarily be combined.

This gives a finite counting prescription.  Denote the accepted
histories by $\mathcal H_{ij}$, using the same vertex-ordering and
symmetry conventions as in the definition of $S_i$.  Then
\[
 m_{ij}=\sum_{H\in\mathcal H_{ij}}1.
\]
A different cyclic spelling of the same intermediate trace is not
a new history, nor is a relabeling of dummy intermediate strings.
The source segment labels are held fixed.  No additional factorial
is inserted for arbitrarily labeling the vertices.  A history records
the placement of reconnections on the graph; an auxiliary order used
to perform operations on independent strings is not counted again.
This describes the long-process counting of \cite{Huang:2010ne} in
terms of the graph.  The two histories
already displayed in Section~2.1 give $m=2$ there.

Why does this discrete count multiply the full field-theory integral?
Fix the segment lengths and impurity positions of any configuration
of type $j$.  An accepted history reconstructs all its intermediate
strings uniquely, simply by performing the prescribed reconnections.
Every intermediate length is a sum of the lengths of its constituent
segments.  Conversely, expanding the cubic vertices into individual
cut choices and summing intermediate position states produces just
such a history.  These constructions are inverse to one another.
The allowed reconnections depend on the cyclic ordering, not on the
positive lengths or on the impurity positions.  Consequently the
number of sewings is the same $m_{ij}$ at every such configuration.
Here $S_i$ includes all allowed impurity routings on its internal
strings, as in its definition; a routing is fixed once the history
and impurity positions are given.

\medskip
\noindent\emph{From sewing variables to the simplex.}
In the mode description, summing over intermediate modes uses the
completeness relation \eqref{tag4}.  It identifies impurity positions
at neighboring vertices up to a common translation around each
intermediate string.  We therefore have internal lengths and relative
translations as sewing variables, just as $x,t,u$ in Section~2.1.
Translations are measured in units of the total incoming length.
If instead an internal string of circumference $\ell$ is described by
a unit-period angle $\theta$, its translation is $\sigma=\ell\theta$
and $\dd\sigma=\ell\,\dd\theta$; this length factor must be retained.
For a vacuum string the translation integral represents its cyclic
factor.  The finite position sums above provide the same measure
without requiring a mode sum on a vacuum string.

Sewing the normalized planar Wick-contraction vertices through a
complete intermediate-state basis reproduces the usual Wick weight,
including cyclic and identical-string factors.  As in the genus-one
calculation, the vertex normalizations combine with Fourier
completeness to give the position measure.  Each specified history
has color factor $N^{-2h}$ and the same external impurity endpoints,
hence the same BMN phase, as its field-theory contraction.

For the change of variables, temporarily distinguish the internal
strings and use their full labeled ranges, retaining the corresponding
symmetry factors.  Trace every vertex cut back along its segment to
the incoming circle.
The order of these cuts can vary across the sewing domain, and formulas
involving cyclic translations can change when a cut crosses a chosen
origin.  Divide the domain according to these choices, retaining the
corresponding labeled history.  Within each region the cyclic orders
and the choices of representatives modulo the internal circumferences
are fixed.  This is the step that replaces the single change of
variables \eqref{tag9} at genus one: at higher genus the change of
variables is generally made separately in several regions.

Choose one cut as the cyclic origin, keeping this marking consistent
on the string and field-theory sides.  Write the remaining incoming
cut positions in order as
\[
 0=z_0<z_1<\cdots<z_{4h-1}<z_{4h}=1.
\]
Define the consecutive segment lengths by
\[
 x_a=z_a-z_{a-1}\quad(1\leq a\leq4h),
 \qquad
 z_a=\sum_{b=1}^a x_b\quad(1\leq a<4h).
\]
These equations give the explicit change of variables.  Its Jacobian
is triangular with diagonal entries $1$, so
\[
 \dd z_1\cdots\dd z_{4h-1}
 =\dd x_1\cdots\dd x_{4h-1},
 \qquad x_{4h}=1-\sum_{a=1}^{4h-1}x_a.
\]
The ordering inequalities become exactly
\[
 \Delta_{4h-1}
 =\left\{\bm x\in\mathbb R_{\geq0}^{4h}:
                   \sum_{a=1}^{4h}x_a=1\right\}.
\]
Coincident cuts are its measure-zero boundary.

It remains to check that the original sewing variables cover the
whole simplex, rather than a smaller region.  Given any positive
$\bm x$ and an accepted history $H$, reconstruct its intermediate
cyclic words.  The circumference of each internal string is the sum
of its segment lengths, and its sewing translation is the sum of the
lengths encountered between the two specified origins on that word.
Thus every sewing variable is recovered uniquely.  Its allowed range
and length conservation follow automatically from positive segment
lengths.  There is no further inequality on $\bm x$.  This is the
inverse map that establishes coverage of the full simplex.

This reconstruction also checks the measure before the final gap
substitution.  On a region with fixed orders and fixed cyclic origins,
both the sewing-to-cut map and its inverse use only additions and
subtractions of lengths and absolute positions.  In independent
coordinates adapted to the finite-$J$ position lattice, both affine
maps have integer linear parts; being inverse, their determinants
are $\pm1$.  Equivalently, the reconstruction is a bijection of the
finite position configurations, so the corresponding Riemann sums
have the same measure.  Fourier completeness, with the vertex
normalization just discussed, gives precisely this position measure.
The remaining
external phases are bounded and piecewise continuous, so these sums
converge to the simplex and impurity-position integrals.

Cyclic markings and diagram symmetry factors must still be treated
with the conventions defining $S_i$ and $F_j$.  A unit Jacobian alone
does not determine those discrete factors.  They are already fixed
by the finite counting above and are carried through the change of
variables.  In particular, one should not count regions in an
arbitrarily labeled sewing parametrization and identify that count
with $m_{ij}$ before removing its extra labels.

For later use, we write the normalized simplex measure as
\begin{equation}
 \dd\mu_h=(4h-1)!\,\dd x_1\cdots\dd x_{4h-1},
 \qquad x_{4h}=1-\sum_{a=1}^{4h-1}x_a,
 \qquad\int_{\Delta_{4h-1}}\dd\mu_h=1.
 \label{eq:normalized-simplex-measure}
\end{equation}
This factorial normalizes the coordinate volume; it is separate from
the normalization of an individual field-theory diagram.

We can now integrate the configuration-by-configuration counting
identity: the field-theory configurations of type $j$ have exactly
$m_{ij}$ allowed histories of shape $i$, with equal weights, phases,
and integration measures.  Hence
\begin{equation}
 S_i=\sum_jm_{ij}F_j.
 \label{eq:general-factorization-proof}
\end{equation}
Thus the multiplicity is supplied by the discrete history count, and
the simplex integral supplies the common continuous contribution.

\medskip
\noindent\emph{The total two-point function.}
To sum the string diagrams, the ordering convention must be kept
explicit.  Let $L_i$ be the number of orderings of the $2h$ cubic
vertices of diagram $i$ compatible with the directions of its
internal strings.  For a fixed field-theory diagram $j$, the $2h$
distinct paired reconnections are disjoint transpositions.  They
commute, so each of their $(2h)!$ orders gives the same final cyclic
word.  Every order determines a string diagram, an accepted assignment
of reconnections to its vertices, and a compatible vertex ordering.
Conversely, an assignment counted by $m_{ij}$ has exactly $L_i$ such
orders.  Operations on independent strings can be interchanged without
changing the graph assignment or its contribution.  Therefore
\begin{equation}
 \sum_i L_i m_{ij}=(2h)!,
 \label{eq:total-factorization-multiplicity}
\end{equation}
independently of the field-diagram index $j$.  Combining this count
with \eqref{eq:general-factorization-proof} gives
\begin{equation}
 \big\langle\overline{O}_{\bm m}O_{\bm n}\big\rangle_h
 =\sum_j F_j
 =\frac{1}{(2h)!}\sum_i L_i S_i.
 \label{eq:total-string-diagram-sum}
\end{equation}
Thus the total genus-$h$ two-point function is a sum of string-diagram
contributions with positive coefficients.

At genus one and two,  every diagram
has $L_i=1$, and the unweighted multiplicity sums are respectively
$2$ and $24$.  This simplification does not hold at higher genus. 
For example at genus three, there are 22 string diagrams and the possible values of  $L_i$ are $\{1,2,3,4,6\}$.
In Appendix \ref{genusthree} we provide the genus three string diagrams and explain how to count the coefficient $L_i$ with some examples.

\section{Non-negativity of  BMN two-point functions}  \label{secnonnegative}

In the strict BMN limit $J, N \rightarrow \infty $, the higher point functions vanish and are regarded as a kind of virtual processes. The finite BMN two-point functions are real and symmetric at each genus.  Define the properly normalized all-genus BMN two-point functions 
\be  \label{definetwopoint}
p_{{\bm m}, {\bm n}} : = \frac{g}{2\sinh (\frac{g}{2})} \sum_{h=0}^{\infty}  \big\langle  \overline{O}_{\bm m} O_{\bm n}\big\rangle_{h }, 
\ee
where the genus counting parameter $g=\frac{J^2}{N}$ is identified as the string coupling constant.  In \cite{Huang:2019lso, Du:2021dml, Du:2021spv}, we propose a probability interpretation that this is holographically identified with the norm square of the quantum unitary transition amplitudes between the corresponding degenerate tensionless strings for the cases of no more than three string modes
\be  \label{probability}
p_{{\bm m}, {\bm n}} = | \langle {\bm m} |\hat{U}(g) | {\bm n}\rangle |^2, ~~~ k=2,3. 
\ee
So by completeness of the orthonormal BMN string basis we confirm the relation $\sum_{\bm n} p_{{\bm m}, {\bm n}} =1$, which can be derived from gauge theory side. 

Here the single-trace BMN operators are normalized by the free planar two point function, and for convenience we always omit the universal spacetime factors in the free BMN correlators. So the time direction of unitary transition on the string theory side is emergent from the perspective of gauge theory side.

A consequence of the proposal is the non-negativity of BMN two-point functions. We should note that by gauge theory definition (\ref{definetwopoint}) and assuming good behaviors in the strict BMN limit, the matrix $P=\{ p_{{\bm m}, {\bm n}} \}$ is a Gram matrix, whose eigenvalues and diagonal elements are always non-negative. It is a different and much stronger statement from holography (\ref{probability}) that all matrix elements are non-negative. The non-negativity of $p_{{\bm m}, {\bm n}}$ is easy to prove at each genus for the case of two string modes. We made the stronger conjecture that this was also true for the case of three string modes separately  at each genus, i.e. 
\be \label{conjecture1}
 \big\langle \overline{O}_{m_1, m_2, m_3} O_{n_1, n_2, n_3}\big\rangle_{h } \geq 0, ~~~ \sum_{i=1}^3 m_i =\sum_{i=1}^3 n_i =0. 
\ee
We proved the non-negativity at genus one and  performed extensive tests at genus two in \cite{Du:2021spv}.  The generic large-mode behavior also supports positivity at higher genus.  On the other hand, for four or more string modes, although the matrix $P$ is still a Gram matrix, some matrix elements can be negative  \cite{Du:2021dml}. 

Usually the computations are  quite complicated. Now with some simple prompts, AI tools can implement the algorithm with some self improvements to perform much more extensive tests.  In the Appendix \ref{appendixcounter} we discuss an even stronger conjecture (\ref{eq:string-diagram-positivity-conjecture}) which would imply (\ref{conjecture1}). The stronger conjecture can be proven at genus two and also passes extensive tests at genus 3, 4, 5. So this proves the conjecture (\ref{conjecture1}) at genus two for all mode numbers and  provides additional tests  at higher genus. However, it turns out the stronger conjecture (\ref{eq:string-diagram-positivity-conjecture}) is ultimately disproved by an ingenious counterexample at large genus, as discussed in the Appendix \ref{appendixcounter}. The main conjecture (\ref{conjecture1}) remains unresolved. 

We perform more tests of the full correlator (\ref{conjecture1}) directly, independently
of the disproved stronger conjecture (\ref{eq:string-diagram-positivity-conjecture}). For simplicity here we just provide  the results of a one-prompt test with running time of  about 15 minutes. At $h=3$, exact integration and
summation over all $1,485$ diagrams rigorously establish positivity for $183$
selected mode pairs. We also use an unbiased Monte Carlo (MC) evaluation,
sampling admissible diagrams uniformly and interval lengths uniformly on
the corresponding simplex, with the impurity positions integrated analytically.
At each $h=3,4,5,6$, the broad scan covers all level-matched pairs with
$0<|m_a|,|n_a|\leq16$ ($a=1,2,3$), modulo simultaneous impurity permutations,
simultaneous sign reversal and exchange of ${\bm m}$ and ${\bm n}$, giving
$21,872$ classes. Table~\ref{tab:total-correlator-positivity-tests} also includes
selected tests at larger modes and genera. Every MC test has a strictly
positive lower confidence bound at simultaneous $99\%$ confidence after
correction for multiple comparisons; these are statistical tests, whereas
the exact genus-three checks certify positivity for the selected pairs. No negative result was found in the tests.

\begin{table}[!ht]
\centering
\small
\begin{tabular}{c l r c}
\hline
Genus $h$ & Evaluation & Mode pairs & Samples per pair \\
\hline
$3$ & Exact, selected modes & $183$ & --- \\
$3$ & MC, broad scan & $21,872$ & $2^{21}$ or $2^{22}$ \\
$4$ & MC, broad scan & $21,872$ & $2^{21}$ or $2^{22}$ \\
$5$ & MC, broad scan & $21,872$ & $2^{21}$ or $2^{22}$ \\
$6$ & MC, broad scan & $21,872$ & $2^{21}$ or $2^{22}$ \\
$6$ & MC, selected larger modes & $36$ & $2^{21}$ or $2^{24}$ \\
$10,20$ & MC, selected modes & $3$ each & $2^{20}$ \\
\hline
\end{tabular}
\caption{Independent tests of (\ref{conjecture1}), all supporting positivity.
The exact genus-three and selected genus-six tests reach
$\max_a\{|m_a|,|n_a|\}=64$ and $576$, respectively.
The mode pairs in different rows can overlap.}
\label{tab:total-correlator-positivity-tests}
\end{table}

In mathematical terminology, a matrix with non-negative elements whose rows sum to one
is called a \emph{row stochastic matrix}; if its columns also sum to one, it is
\emph{doubly stochastic}, or \emph{bistochastic}. A matrix satisfying
\eqref{probability}, with entries equal to the squared moduli of a unitary
matrix, is called \emph{unistochastic}. Such matrices have been studied,
for example, in \cite{Bengtsson:2004unistochastic}.
Every unistochastic matrix is doubly stochastic, but the converse is
false: row and column normalization do not ensure that phases can be
chosen to make all rows mutually orthogonal. In our case, symmetry and
normalization imply that $P$ is doubly stochastic  assuming non-negativity in the case of three string modes, 
whereas \eqref{probability} imposes further conditions, which in our context look much weaker than the 
non-negativity.
Appendix~\ref{appendixunistochastic} explains the necessary polygon-closure
conditions and gives a common small-coupling proof for each fixed pair
of rows with two or three string modes.

\section{Entropy of BMN strings}

We start from an initial BMN state $|{\bm m}\rangle$ and go through a unitary evolution with string coupling $g$, then measure in the BMN basis.  We define the entropy of BMN strings in \cite{Huang:2019uue} as the von Neumann entropy of the resulting mixed state.  In this section for simplicity we consider the case of two string modes, and denote the two-point functions as
\[
 A_h(m,n):=\left\langle \overline{O}_{-m,m}^{J}O_{-n,n}^{J}\right\rangle_h,
 \qquad
 p_{m,n}(g)=\frac{g}{2\sinh(g/2)}\sum_{h\geq 0}A_h(m,n),
\]
and the entropy as 
\[
 S_m(g):=-\sum_{n\in\mathbb Z}p_{m,n}(g)\log p_{m,n}(g).
\]
 For the large-mode lower bound we will use the Gram-matrix property of the normalized
all-genus correlator.  

We propose a cosmological application of the BMN strings in \cite{Huang:2025dnq}. To test the idea, it is important to understand the behavior of the entropy, especially in the strong coupling limit $g\rightarrow \infty$. In this section we provide several mathematical results on this issue.

\subsection{An improved upper bound}

We proved an upper bound for the entropy $S_m(g) < (2+\epsilon)\log(g) + o(1)$ in the strong coupling limit $g\rightarrow\infty$ in \cite{Huang:2019uue}. Here we improve the logarithmic coefficient from $2+\epsilon$ to $1$. The argument uses the Fourier representation of each diagram, Parseval's identity, and a bound on the number of interval boundaries averaged over the genus weights.

For a genus-$h$ diagram $D$ with interval lengths
$x_1,\ldots,x_{4h}$, let $P_{D,\bm x}$ be the associated measure-preserving
piecewise translation of the circle and let $Q_{D,\bm x}=P_{D,\bm x}^{-1}$.
For a transition from $m$ to $n=m+k$, the one-mode factor in the diagram
integral has the form
\begin{align}
 I_{D,\bm x}(m,n)
 &=\int_0^1\exp\!\left(2\pi i\,[nP_{D,\bm x}(y)-my]\right)\,\dd y \\
 &=\int_0^1F_{m,D,\bm x}(z)e^{2\pi i k z}\,\dd z,
 \qquad
 F_{m,D,\bm x}(z):=
 \exp\!\left(2\pi i m[z-Q_{D,\bm x}(z)]\right).
 \label{eq:mode-dependent-phase}
\end{align}
Here the second line follows from the measure-preserving change of variable
$z=P_{D,\bm x}(y)$. Thus the Fourier function is $F_{m,D,\bm x}$, which
depends on the initial mode $m$ but not on the difference $k=n-m$, and
$|F_{m,D,\bm x}|=1$. With
\[
 \widehat F(k):=\int_0^1F(z)e^{-2\pi i k z}\,\dd z,
\]
the absolute square of the factor in \eqref{eq:mode-dependent-phase} is
$|\widehat F_{m,D,\bm x}(-k)|^2$.

Set
\begin{equation}
 Z(g):=\frac{2\sinh(g/2)}{g},
 \qquad
 w_h(g):=\frac{(g/2)^{2h}}{(2h+1)!Z(g)},
 \qquad \sum_{h\geq0}w_h(g)=1.
 \label{eq:Z-definition}
\end{equation}
Let $\mathcal D_h$ denote the genus-$h$ diagram set and let
$N_h=|\mathcal D_h|$. All diagrams have equal normalized weight $1/N_h$.
With the normalized simplex measure defined in
\eqref{eq:normalized-simplex-measure}, the two-mode diagram representation is
\begin{equation}
 p_{m,m+k}(g)
 =w_0(g)\,\delta_{k0}
 +\sum_{h\geq1}\frac{w_h(g)}{N_h}\sum_{D\in\mathcal D_h}
 \int_{\Delta_{4h-1}}
 \left|\widehat F_{m,D,\bm x}(-k)\right|^2\,\dd\mu_h.
 \label{eq:fourier-probability-representation}
\end{equation}
This normalization gives $\sum_kp_{m,m+k}(g)=1$ by Parseval.

For $t\in\mathbb R$, define
\begin{equation}
 D_F(t):=\int_0^1|F(y+t)-F(y)|^2\,\dd y,
 \label{eq:DF-definition}
\end{equation}
where $F$ is understood periodically. Parseval gives
\begin{equation}
 D_F(t)=2\sum_{k\in\mathbb Z}
 \bigl(1-\cos(2\pi kt)\bigr)|\widehat F(k)|^2.
 \label{eq:parseval}
\end{equation}
The function $F_{m,D,\bm x}$ is constant on each of the $4h$ image
intervals. For $0\leq t\leq1$, it can change only when the arc from $y$
to $y+t$ crosses one of the $4h$ boundaries. The set of such $y$ has
measure at most $4ht$, and $|F(y+t)-F(y)|^2\leq4$. Consequently,
\begin{equation}
 D_{F_{m,D,\bm x}}(t)\leq16ht,
 \qquad t\geq0.
 \label{eq:boundary-bound}
\end{equation}
For $t>1$ the same bound follows from $D_F(t)\leq4$ and $h\geq1$.
It is independent of the initial mode, the interval lengths, and the
diagram ordering.

The averaged translation quantity is
\begin{equation}
 D_g^{(m)}(t):=
 \sum_{h\geq1}\frac{w_h(g)}{N_h}\sum_{D\in\mathcal D_h}
 \int_{\Delta_{4h-1}}D_{F_{m,D,\bm x}}(t)\,\dd\mu_h;
 \label{eq:Dg-definition}
\end{equation}
the planar contribution is zero. Differentiating the series for $Z(g)$ gives
\[
 \sum_{h\geq0}4h\,w_h(g)
 =\frac{2gZ'(g)}{Z(g)}
 =g\coth\frac g2-2
 =g+\frac{2g}{e^g-1}-2
 \leq g,
\]
where the last inequality uses $e^g-1\geq g$ for $g>0$.
Thus \eqref{eq:boundary-bound} gives the uniform estimate
\begin{equation}
 D_g^{(m)}(t)\leq4gt,
 \qquad g>0,\quad t\geq0.
 \label{eq:Dg-global-bound}
\end{equation}

For $a>0$, introduce the comparison distribution
\[
 q_a(k):=\frac{(1+k^2/a^2)^{-1}}{\mathcal Z(a)},
 \qquad
 \mathcal Z(a):=\sum_{k\in\mathbb Z}(1+k^2/a^2)^{-1}.
\]
Non-negativity of relative entropy gives
\begin{equation}
 S_m(g)\leq\log\mathcal Z(a)
 +\sum_{k\in\mathbb Z}p_{m,m+k}(g)
 \log\left(1+\frac{k^2}{a^2}\right).
 \label{eq:gibbs-comparison}
\end{equation}
The elementary Fourier identity
\begin{equation}
 \log\left(1+\frac{k^2}{a^2}\right)
 =2\int_0^\infty e^{-2\pi at}
 \frac{1-\cos(2\pi kt)}{t}\,\dd t
 \label{eq:log-fourier}
\end{equation}
and Parseval, with Tonelli's theorem for the nonnegative integrand, give
\begin{equation}
 \sum_{k\in\mathbb Z}p_{m,m+k}(g)
 \log\left(1+\frac{k^2}{a^2}\right)
 =\int_0^\infty e^{-2\pi at}\frac{D_g^{(m)}(t)}{t}\,\dd t
 \leq\frac{2g}{\pi a}.
 \label{eq:log-moment}
\end{equation}
Since $(1+x^2/a^2)^{-1}$ decreases for $x\geq0$, comparison with its
integral also gives
\[
 \mathcal Z(a)\leq1+2\int_0^\infty\frac{\dd x}{1+x^2/a^2}
 =1+\pi a.
\]
Taking $a=2g/\pi$ in \eqref{eq:gibbs-comparison}, we obtain
\begin{equation}
 S_m(g)\leq\log(1+2g)+1
 =\log g+O(1),\qquad g\to\infty.
 \label{eq:general-upper-bound}
\end{equation}
The explicit bound holds for every $g>0$ and every initial mode $m$;
in particular, the $O(1)$ constant is independent of $m$.

\subsection{Lower bounds}

\label{subsecunbounded}

We now work in the opposite direction and derive lower bounds for the entropy.  First we derive a diagonal inequality, which follows from the real symmetric and 
Gram property and row normalization $\sum_n p_{m,n}=1$.  Let $P(g)=(p_{m,n}(g))$ act on
$\ell^2(\mathbb Z)$.  For $ v \in \ell^2(\mathbb Z)$, by the weighted Cauchy-Schwarz inequality,
\begin{align*}
 |(Pv)_m|^2
 = |\sum_n p_{m,n}v_n |^2
 \leq
 (\sum_n p_{m,n} )
 (\sum_n p_{m,n}|v_n|^2 )
 =\sum_n p_{m,n}|v_n|^2.
\end{align*}
Summing over $m$  gives $\|Pv\|^2 \leq  \|v\|^2$, hence $0\leq  P \leq I$ and
$P  -P ^2 = P^{\frac{1}{2}} (I-P)  P^{\frac{1}{2}}  \geq 0 $.  Taking the $m$-th diagonal entry yields
\begin{equation}
 \sum_n p_{m,n}(g)^2=(P(g)^2)_{m,m}\leq p_{m,m}(g).
 \label{eq:entropy-diagonal-bound}
\end{equation}
Since \(\log\) is concave, Jensen's inequality implies that the order-two Rényi entropy is no larger than the Shannon entropy, therefore
\begin{equation}
 S_m(g)\geq
 -\log\sum_n p_{m,n}(g)^2\geq-\log p_{m,m}(g).
 \label{eq:entropy-diagonal-lower-bound}
\end{equation}
So to get a lower bound for the entropy, we need to establish an upper bound for the diagonal element $p_{m,m}$.

\subsubsection{Large initial mode} 

The Riemann--Lebesgue lemma will be important in the estimate with a large initial string mode. It removes
nonzero oscillatory phases at every fixed genus, but it does not remove cross
terms whose translation functions are identically equal. 

Fix a genus-$h$ diagram $D$ with $h\geq1$.  Label its $4h$ intervals by
$i=1,\ldots,4h$, and let $\Delta_{D,i}$ be the translation function on the
$i$th interval.  Define the translation classes
\begin{equation}
 i\sim_D j
 \quad\Longleftrightarrow\quad
 \Delta_{D,i}\equiv\Delta_{D,j},
 \qquad
 {\cal C}(D):=\{\,\text{equivalence classes}\,\},
 \qquad
 \ell_C:=|C|.
 \label{eq:translation-classes}
\end{equation}
The class $C$ can contain distinct intervals.

We define $r_{h,D}(m)$ to be the normalized diagonal contribution of diagram
$D$ and set
\begin{equation}
 r_h(m):=\frac{1}{N_h}\sum_{D\in\mathcal D_h}r_{h,D}(m),\qquad 0\leq r_h(m)\leq1.
 \label{eq:rh-definition}
\end{equation}
For the planar identity contribution we set $r_0(m)=1$ (and the
corresponding off-diagonal planar entries to zero).  The relation to the
probability is then fixed by
\begin{equation}
 p_{m,m}(g)=\sum_{h\geq0}w_h(g)r_h(m),\qquad w_h(g)=\frac{(g/2)^{2h}}{(2h+1)!Z(g)}.
 \label{eq:rh-probability-mixture}
\end{equation}
For the diagonal matrix element $n=m$, the $m$-dependent part of a fixed
diagram is a finite sum of terms
\[
 \sum_{i=1}^{4h}x_i e^{2\pi i m\Delta_{D,i}}.
\]
After squaring, a term with indices $(i,j)$ has phase
$e^{2\pi i m(\Delta_{D,i}-\Delta_{D,j})}$.  In the $m\rightarrow \infty$ limit,  the Riemann--Lebesgue lemma, or
the elementary estimate 
\begin{equation}
 \int_a^b e^{2\pi i m c y}\,\dd y
 =\frac{e^{2\pi i m c b}-e^{2\pi i m c a}}{2\pi i m c}
 \longrightarrow0
 \qquad(c\neq0),
 \label{eq:elementary-RL-step}
\end{equation}
kills the terms for which $\Delta_{D,i}-\Delta_{D,j}$ is not identically zero
only after a multidimensional Fubini/Riemann--Lebesgue argument on the
simplex.  The terms with $i\sim_D j$ survive, and the fixed-genus large-mode
limit for this diagram is
\begin{equation}
 \lim_{m\to\infty}r_{h,D}(m)
 =\int_{\Delta_{4h-1}}
 \sum_{C\in{\cal C}(D)}
 \left(\sum_{i\in C}x_i\right)^2 \dd\mu_h .
 \label{eq:class-sum-before-simplex}
\end{equation}

Using the normalized simplex measure \eqref{eq:normalized-simplex-measure},
the Dirichlet simplex moments for a class of size $\ell_C$ give
\begin{equation}
 \int_{\Delta_{4h-1}}
 \left(\sum_{i\in C}x_i\right)^2\dd\mu_h
 =\frac{\ell_C(\ell_C+1)}{(4h)(4h+1)}.
 \label{eq:class-simplex-moment}
\end{equation}
Indeed, $\int x_i^2\dd\mu_h=2/[(4h)(4h+1)]$ and
$\int x_i x_j\dd\mu_h=1/[(4h)(4h+1)]$ for $i\neq j$.  Thus the normalized
large-mode limit at genus $h$ is
\begin{equation}
 r_h^{(\infty)}
 =
 \frac{1}{|\mathcal D_h|}
 \sum_{D\in\mathcal D_h}
 \frac{\displaystyle\sum_{C\in{\cal C}(D)}
       \ell_C(\ell_C+1)}
      {(4h)(4h+1)},
 \label{eq:correct-large-m-limit}
\end{equation}
where $\mathcal D_h$ is the set of genus-$h$ diagrams. For each diagram $D$ there are $4h$ segments, we have $\sum_{C\in{\cal C}(D)} \ell_C = 4h$. 

If every class were a singleton, \eqref{eq:correct-large-m-limit} would reduce
to
\[
 r_h^{(\infty)}=\frac{2}{4h+1}.
\]
With non-singleton classes there is an additional positive term:
\begin{equation}
 r_h^{(\infty)}
 =
 \frac{2}{4h+1}
 +
 \frac{1}{(4h)(4h+1)|\mathcal D_h|}
 \sum_{D\in\mathcal D_h}
 \sum_{C\in{\cal C}(D)}\ell_C(\ell_C-1).
 \label{eq:cross-class-correction}
\end{equation}

The variable-mode unboundedness argument requires a bound on the cross terms.
Here the required bound follows from the marked-pair counting identity.  Let
\begin{equation}
 N_h:=|\mathcal D_h|=\frac{(4h-1)!!}{2h+1}\quad(h\geq1),
 \qquad N_1=1,
 \label{eq:number-of-diagrams}
\end{equation}
be the number of genus-$h$ diagrams with a fixed source origin. This count follows from the pairing enumeration of Harer and Zagier~\cite{Harer:1986}. For fixed $h$,
let $n_s$ be the total number of translation classes of size $s$, counted
across all diagrams in $\mathcal D_h$, so we have $\sum_s sn_s =4hN_h$.  Denote 
\begin{equation}
 Q_h: = \sum_{D\in\mathcal D_h}
 \sum_{C\in{\cal C}(D)}\ell_C(\ell_C-1) = \sum_{s\geq2}s(s-1)n_s.
 \label{eq:class-size-count}
\end{equation}
The marked-pair counting identity is
\begin{equation}
 \boxed{\qquad
 Q_h=4h\sum_{r=1}^{h-1}N_rN_{h-r}.
 \qquad}
 \label{eq:marked-pair-convolution}
\end{equation}
For $h=1$ the sum is empty, so $Q_1=0$.  Exhaustive enumeration gives
\begin{equation}
 (Q_1,Q_2,Q_3,Q_4,Q_5)
 =(0,8,504,54576,10256400),
 \label{eq:marked-pair-checks}
\end{equation}
which agrees with \eqref{eq:marked-pair-convolution} using
$(N_1,N_2,N_3,N_4,N_5)=(1,21,1485,225225,59520825)$.

To our knowledge, the combinatorial identity (\ref{eq:marked-pair-convolution}) is new. It was identified with AI assistance from low-genus enumeration. To prove it for all genera, count diagrams
$D\in\mathcal D_h$ with an ordered pair $(i,j)$ of distinct intervals such
that $\Delta_{D,i}\equiv\Delta_{D,j}$ as functions of the interval lengths.
Since a translation is the difference between an interval's starting
positions on the source and target circles, this equality means that the
arcs between the marked starting points contain exactly the same interval
labels on both circles, possibly in different orders.  In the usual
boundary-pairing description, the boundaries at the ends of these intervals
therefore have their partners within the same group.  Cutting at the two
marked positions and closing each part separately consequently produces two
allowed diagrams.  Each part has a single target circle: any additional
closed component would already have been a separate component of the
original diagram.  The two pieces satisfy the same pairing rule as the
original diagrams and have respectively $4r$ and $4(h-r)$ intervals, with
$1\leq r\leq h-1$.  Conversely, joining any two such diagrams at their marked
starting points uniquely reconstructs the original diagram and its ordered
pair of equal translations.  For a fixed first mark and a fixed $r$, there
are therefore $N_rN_{h-r}$ possibilities; the second mark lies $4r$ intervals
after the first.  There are $4h$ choices for the first mark, and a translation
class $C\in{\cal C}(D)$ contributes $\ell_C(\ell_C-1)$ ordered pairs.  Thus
$Q_h=\sum_{D\in\mathcal D_h}\sum_{C\in{\cal C}(D)}\ell_C(\ell_C-1)
=4h\sum_{r=1}^{h-1}N_rN_{h-r}$, proving
\eqref{eq:marked-pair-convolution}.  No additional symmetry factor is
required because the source intervals are labeled and the two marks fix the
cutting and joining positions.

We now turn this identity into the estimate needed below.  The ratios of the
diagram numbers are
\begin{equation}
 R_k:= \frac{N_{k+1}}{N_k}
 =
 \frac{(4k+1)(4k+3)(2k+1)}{2k+3}
 =
 16k^2+11-\frac{30}{2k+3},
 \label{eq:diagram-ratio}
\end{equation}
and are strictly increasing for $k\geq1$, since
\[
 R_{k+1} - R_k 
 =
 32k+16+\frac{60}{(2k+3)(2k+5)}>0.
\]
It follows that
\begin{equation}
 N_rN_{h-r}\leq N_{h-1}
 \qquad(1\leq r\leq h-1),
 \label{eq:convolution-term-bound}
\end{equation}
because, writing $b=h-r$, one has
$N_{h-1}/N_r=\prod_{t=0}^{b-2}R_{r+t}$, whereas
$N_b/N_1=\prod_{t=0}^{b-2}R_{1+t}$.  Since $r\geq1$ and the ratios
$R_k $ increase, every factor in the first product is at least the
corresponding factor in the second; $N_1=1$ then gives
$N_{h-1}/N_r\geq N_b$.  Hence
\begin{equation}
 \frac{Q_h}{N_h}
 \leq
 4h(h-1)\frac{N_{h-1}}{N_h}
 =
 \frac{4h(h-1)(2h+1)}
 {(2h-1)(4h-1)(4h-3)}
 \leq1.
 \label{eq:class-count-estimate}
\end{equation}
Equation
\eqref{eq:class-count-estimate} follows from the proved marked-pair identity
\eqref{eq:marked-pair-convolution} and the diagram count. 

Using \eqref{eq:class-count-estimate} and 
\eqref{eq:cross-class-correction} gives
\begin{equation}
 r_h^{(\infty)}
 \leq\frac{8h+1}{(4h)(4h+1)}
 \leq\frac{9}{16h}.
 \label{eq:large-mode-diagonal-bound}
\end{equation}
Using \eqref{eq:class-count-estimate}, choose for each $H$ a threshold $M_H$
such that
\begin{equation}
 r_h(m)\leq\frac{C_1}{h}
 \qquad
 (1\leq h\leq H,\ m\geq M_H),
 \label{eq:finite-genus-large-mode-bound}
\end{equation}
where we choose $C_1=5/8>9/16$.  The explicit rational expression
before the last inequality is strictly smaller than $9/(16h)$ for every
finite $h$, so this choice is compatible with pointwise convergence for each
fixed finite set of genera.  Set
\begin{equation}
 m(g):=\max_{1\leq h\leq\lfloor g\rfloor}M_h
 \qquad (g\geq1).
 \label{eq:choice-of-growing-mode}
\end{equation}
Then, using $0\leq r_h(m)\leq1$ for the remaining genera,
\begin{equation}
 p_{m(g),m(g)}(g)
 \leq w_0(g)
 +C_1\sum_{1\leq h\leq g}\frac{w_h(g)}{h}
 +\sum_{h>g}w_h(g).
 \label{eq:diagonal-split}
\end{equation}
The genus weights have exponentially small tails below a fixed fraction of
$g$ and above $g$, while
$\sum_{h\geq g/8}w_h(g)/h\leq 8/g$.  Hence
\begin{equation}
 p_{m(g),m(g)}(g)\leq\frac{C}{g}+Ce^{-c g}
 \label{eq:diagonal-O-inverse-g}
\end{equation}
for constants $C,c>0$. Therefore, using (\ref{eq:entropy-diagonal-lower-bound}), we see that as $g\rightarrow \infty$, 
\begin{equation}
 S_{m(g)}(g)\geq\log g-O(1)
 \longrightarrow\infty.
 \label{eq:variable-mode-unboundedness}
\end{equation}

\subsubsection{A fixed initial mode}

We now fix an integer $m\neq0$. We will prove the lower bound
\begin{equation}
 S_m(g)\geq\frac14\log g-\frac14\log\log g-O_m(1)
 \longrightarrow\infty,
 \label{eq:fixed-mode-entropy-bound}
\end{equation}
which is weaker than \eqref{eq:variable-mode-unboundedness}, but applies
with the same initial mode at every coupling. By
\eqref{eq:entropy-diagonal-lower-bound}, it suffices to show that
$p_{m,m}(g)=O_m((\log g/g)^{1/4})$.

\medskip
\noindent\emph{Diagram averages and perfect matchings.}
For $h\geq1$, label the source intervals consecutively by
$1,\ldots,4h$. Let $\sigma_D(i)$ be the position of interval $i$ in the
output order, choosing interval $1$ as the origin on both circles, so
that $\sigma_D(1)=1$. The translations introduced above are
\begin{equation}
 \Delta_{D,i}(\bm x)
 =\sum_{\sigma_D(j)<\sigma_D(i)}x_j-\sum_{j<i}x_j
 =\sum_{j=1}^{4h}c_{D,ij}x_j,
 \qquad c_{D,ij}\in\{-1,0,1\},\quad c_{D,ii}=0.
 \label{eq:fixed-mode-translations}
\end{equation}
Write $C_{D,i}=\sum_jc_{D,ij}=\sigma_D(i)-i$.
The diagonal amplitude in \eqref{eq:mode-dependent-phase} and its
normalized genus average are therefore
\begin{align}
 I_{D,\bm x}(m,m)
 &=\sum_{i=1}^{4h}x_i e^{2\pi i m\Delta_{D,i}(\bm x)},
 \nonumber\\
 r_h(m)&=\frac1{N_h}\sum_{D\in\mathcal D_h}
 \int_{\Delta_{4h-1}}|I_{D,\bm x}(m,m)|^2\,\dd\mu_h.
 \label{eq:fixed-mode-diagram-average}
\end{align}
Here $\dd\mu_h$ is the normalized simplex measure in
\eqref{eq:normalized-simplex-measure}; the second identity follows from
\eqref{eq:fourier-probability-representation} at $k=0$.

We use the boundary-pairing description of the leading diagrams.
Let $\gamma=(1\,2\,\cdots\,4h)$ and let $\Omega_h$ be the set of
all perfect matchings of these labels. A perfect matching pairs every
label with exactly one other label, and is represented by a permutation
$\rho$ consisting of $2h$ disjoint transpositions. Put
\begin{equation}
 \beta=\gamma\rho,\qquad
 \mathcal E_h=\{\rho\in\Omega_h:\beta\text{ is a single }4h\text{-cycle}\}.
 \label{eq:fixed-mode-matching-set}
\end{equation}
The cycle condition means that successive applications of $\beta$ visit
every label before returning to the starting point; it expresses that
the output has one trace. For a diagram, its output successor is
$\beta=\sigma_D^{-1}\gamma\sigma_D$, and its boundary pairing is
$\rho=\gamma^{-1}\beta$. The leading $4h$-cut condition makes this
pairing a product of $2h$ transpositions, as in the face-permutation
argument of Section~2. Conversely, a matching in $\mathcal E_h$
determines a unique output order by
$\sigma_D(\beta^j(1))=j+1$ for $0\leq j<4h$.
The absence of fixed points of $\rho$ prevents adjacent intervals from
merging, and its $2h$ transpositions give genus $h$ by Euler's formula.
Thus $\mathcal E_h$ is in bijection with $\mathcal D_h$. By
\eqref{eq:number-of-diagrams},
\begin{equation}
 |\Omega_h|=(4h-1)!!,\qquad |\mathcal E_h|=N_h,
 \qquad \frac{|\mathcal E_h|}{|\Omega_h|}=\frac1{2h+1}.
 \label{eq:fixed-mode-matching-fraction}
\end{equation}
The labels have a fixed origin, so there is no additional quotient by
rotations. For example, at genus two there are $105$ perfect matchings,
of which $21$ give diagrams. A uniform matching average below simply
means a finite sum over $\Omega_h$ divided by $(4h-1)!!$; the diagram
average is the corresponding sum over $\mathcal E_h$ divided by $N_h$.
No random choice of the diagram set is assumed.

\medskip
\noindent\emph{Comparison with equal interval lengths.}
Define
\begin{equation}
 B_D(m)=\frac1{4h}\sum_{i=1}^{4h}
 e^{2\pi i m C_{D,i}/(4h)}.
 \label{eq:fixed-mode-equal-length-amplitude}
\end{equation}
This is the amplitude at $x_i=1/(4h)$. Both $|I_{D,\bm x}(m,m)|$
and $|B_D(m)|$ are at most one: they are averages of unit-modulus
phases with nonnegative weights summing to one, so the triangle
inequality applies. We claim the following uniform comparison:
\begin{equation}
 \int_{\Delta_{4h-1}}|I_{D,\bm x}(m,m)-B_D(m)|^2\,\dd\mu_h
 \leq\frac{8\pi^2m^2+2}{4h}.
 \label{eq:fixed-mode-length-error}
\end{equation}
To verify it, put $T_{D,i}=\sum_jc_{D,ij}^2\leq4h-1$.
The elementary simplex moments, using $c_{D,ii}=0$, give
\begin{align*}
 \int x_i\,\dd\mu_h&=\frac1{4h},\qquad
 \int x_i\Delta_{D,i}\,\dd\mu_h
 =\frac{C_{D,i}}{(4h)(4h+1)},\\
 \int x_i\Delta_{D,i}^2\,\dd\mu_h
 &=\frac{C_{D,i}^2+T_{D,i}}{(4h)(4h+1)(4h+2)}.
\end{align*}
Consequently,
\begin{align}
 \int x_i\left(\Delta_{D,i}-\frac{C_{D,i}}{4h}\right)^2\dd\mu_h
 &=\frac{T_{D,i}}{(4h)(4h+1)(4h+2)}
 +\frac{(2-4h)C_{D,i}^2}{(4h)^3(4h+1)(4h+2)}
 \nonumber\\
 &\leq\frac1{(4h+1)(4h+2)}.
 \label{eq:fixed-mode-simplex-error}
\end{align}
All integrals here and below are over $\Delta_{4h-1}$ when the domain
is omitted. Set $a_i=e^{2\pi i m C_{D,i}/(4h)}$ and
$J_D(\bm x)=\sum_i x_i a_i$. Weighted Cauchy--Schwarz and
$|e^{is}-e^{it}|\leq|s-t|$ imply
\[
 |I_{D,\bm x}(m,m)-J_D(\bm x)|^2
 \leq4\pi^2m^2\sum_i x_i
 \left(\Delta_{D,i}-\frac{C_{D,i}}{4h}\right)^2.
\]
Its integral is at most $4\pi^2m^2/(4h)$ by
\eqref{eq:fixed-mode-simplex-error}. The second simplex moments also give
\[
 \int|J_D(\bm x)-B_D(m)|^2\,\dd\mu_h
 =\frac{1-|B_D(m)|^2}{4h+1}\leq\frac1{4h+1}.
\]
Combining these two estimates with
$|z+w|^2\leq2|z|^2+2|w|^2$ proves
\eqref{eq:fixed-mode-length-error}.

\medskip
\noindent\emph{Short orbits and their full matching average.}
For the rest of the fixed-genus argument assume $4h>|m|$.
On $\mathbb C^{4h}$ use the usual Hermitian inner product and define
\[
 u(a)=\frac{e^{2\pi i ma/(4h)}}{\sqrt{4h}},\qquad
 v_D(a)=\frac{e^{2\pi i m\sigma_D(a)/(4h)}}{\sqrt{4h}},\qquad
 (U_\beta f)(a)=f(\beta(a)).
\]
The operator $U_\beta$ is unitary. On $\mathcal E_h$,
$U_\beta v_D=\lambda v_D$ with $\lambda=e^{2\pi i m/(4h)}$,
and $B_D(m)=\langle u,v_D\rangle$. For a positive integer $K$, let
\begin{equation}
 T_K=\frac1K\sum_{j=0}^{K-1}\lambda^{-j}U_\beta^j,
 \qquad
 f_\ell(\rho)=\langle u,U_\beta^\ell u\rangle
 =\frac1{4h}\sum_{a=1}^{4h}
 e^{2\pi i m(\beta^\ell(a)-a)/(4h)}.
 \label{eq:fixed-mode-orbit-correlation}
\end{equation}
Since $T_Kv_D=T_K^*v_D=v_D$, Cauchy--Schwarz and expansion of
$\|T_Ku\|^2$ yield
\begin{equation}
 |B_D(m)|^2\leq\|T_Ku\|^2
 \leq\frac1K+\frac2{K^2}
 \sum_{\ell=1}^{K-1}(K-\ell)|f_\ell(\rho)|.
 \label{eq:fixed-mode-orbit-reduction}
\end{equation}
Each summand defining $f_\ell$ is a unit-modulus phase; $f_\ell$ itself
is their average, so $|f_\ell|\leq1$. We first estimate its full
matching average
\[
 \bar f_\ell=\frac1{|\Omega_h|}\sum_{\rho\in\Omega_h}f_\ell(\rho),
\]
and will then control the average of $|f_\ell|$ on $\mathcal E_h$.
These are different quantities: cancellation in $\bar f_\ell$ alone
does not bound the average absolute value on a subset.

Fix a starting label $a$ and follow $z_t=\beta^t(a)$, revealing the
pair containing the current label only when needed. Call a step fresh
if its input has not appeared in a previously revealed pair. Before
the $t$th fresh step, $t-1$ pairs have been exposed, so the current
label has $4h-2t+1$ possible partners. Each partner choice has exactly
the same number of complete matchings extending the revealed pairs.
If the partner is $b$, the next label is $\gamma(b)$. For this next
label to be already paired, it must lie among the $2t-2$ previously
exposed labels or equal the current label; it cannot equal $b$,
since $\gamma$ has no fixed point. There are therefore at most $2t-1$
partner choices causing this collision, and some may already be
unavailable.

Let $\mathcal B_{a,\ell}\subseteq\Omega_h$ be the set of matchings
for which at least one of the first $\ell$ steps is not fresh. Counting
according to the first collision gives, for $1\leq\ell\leq h$,
\begin{equation}
 \frac{|\mathcal B_{a,\ell}|}{|\Omega_h|}
 \leq\sum_{t=1}^{\ell-1}\frac{2t-1}{4h-2t+1}
 \leq\sum_{t=1}^{\ell-1}\frac{2t}{4h-2t+1}
 \leq\frac{\ell(\ell-1)}{4h-2\ell+1}.
 \label{eq:fixed-mode-collision-fraction}
\end{equation}
This is an upper bound, not an exact formula for the bad fraction.
The displayed denominators count partners after a fresh history,
whereas the bad fraction on the left counts full matchings in
$\Omega_h$. The bound holds for every starting label $a$.

For any history in which the first $\ell-1$ steps and the next input
are all fresh, the last endpoint ranges equally over $4h-2\ell+1$ labels.
Because $m\neq0$ and $|m|<4h$, the exact Fourier identity is
\[
 \sum_{b=1}^{4h}e^{2\pi i mb/(4h)}=0.
\]
Removing $2\ell-1$ terms from this sum leaves a sum of modulus at most
$2\ell-1$. Thus the average last phase for each such history has
modulus at most $(2\ell-1)/(4h-2\ell+1)$. The other matchings
contribute at most their fraction, since each phase has modulus one.
Multiplying by the starting phase and then averaging over all $4h$
starting labels gives
\begin{equation}
 |\bar f_\ell|
 \leq\frac{\ell^2+\ell-1}{4h-2\ell+1}
 \leq\frac{6\ell^2}{4h},\qquad 1\leq\ell\leq h.
 \label{eq:fixed-mode-full-mean}
\end{equation}
The factor $1/(4h)$ in $f_\ell$ cancels the number of starting labels;
there is no further factor $4h$ in this estimate.

\medskip
\noindent\emph{A finite-average form of Azuma's inequality.}
Switching two pairs,
$\{a,b\},\{c,d\}\leftrightarrow\{a,c\},\{b,d\}$, changes $\beta$
on at most four inputs. The two $\ell$th powers can differ only for
starting labels whose orbit meets one of these inputs during the first
$\ell$ steps, so they differ on at most $4\ell$ inputs. Each summand
in $f_\ell$ changes by at most two. A switch therefore changes
$f_\ell$ by at most
\[
 L_\ell=\frac{8\ell}{4h}.
\]
This bounds one local change of a matching; it does not bound the
difference between an arbitrary matching and the full average by
$L_\ell$.

To control that difference, reveal the partner of the smallest unmatched
label successively. For either the real or the imaginary part
$F$ of $f_\ell$, let $M_j$ be its average over all complete matchings
agreeing with the first $j$ revealed pairs. Thus $M_0$ is the full
average with no pairs specified, while $M_{2h}=F(\rho)$ is the value
for one complete matching. Every possible next partner has the same
number of completions, so each parent average is the ordinary average
of its child averages. Furthermore, completions for two partner
choices are in bijection by the two-pair switch above. Their child
averages differ by at most $L_\ell$, and hence
$|M_j-M_{j-1}|\leq L_\ell$.

For completeness, the concentration estimate can be proved directly
with these finite averages. If $d_b$ is a child average minus its
parent average, then its average over the possible partners $b$ is
zero and $|d_b|\leq L_\ell$. Convexity gives, for real $s$,
\[
 \operatorname{avg}_b e^{s d_b}
 \leq\cosh(sL_\ell)\leq e^{s^2L_\ell^2/2}.
\]
Iterating through the $2h$ pairs, with
$\bar F=|\Omega_h|^{-1}\sum_{\rho\in\Omega_h}F(\rho)$, gives
\[
 \frac1{|\Omega_h|}\sum_{\rho\in\Omega_h}
 e^{s(F(\rho)-\bar F)}\leq e^{h s^2L_\ell^2}.
\]
For $t>0$, the fraction with $F-\bar F>t$ is at most
$e^{-st+h s^2L_\ell^2}$; minimizing over $s>0$ bounds it by
$e^{-t^2/(4hL_\ell^2)}$. Apply this to both signs of the real and
imaginary parts, observing that a complex number of modulus greater
than $t$ has one component of absolute value greater than $t/\sqrt2$.
We obtain
\begin{equation}
 \frac{\#\{\rho\in\Omega_h:|f_\ell(\rho)-\bar f_\ell|>t\}}
 {|\Omega_h|}
 \leq4\exp\!\left(-\frac{(4h)t^2}{128\ell^2}\right).
 \label{eq:fixed-mode-concentration}
\end{equation}
This is Azuma's inequality~\cite{Azuma:1967} expressed as a bound on a fraction of a
finite set. It requires no independence of the successive revealed
pairs: the equal-completion averages and the switch bound supply the
needed hypotheses.

Take $t_\ell=16\ell\sqrt{\log(4h)/(4h)}$ and denote the exceptional
set in \eqref{eq:fixed-mode-concentration} by $\mathcal X_\ell$.
Its fraction in $\Omega_h$ is at most $4/(4h)^2$. Using
\eqref{eq:fixed-mode-matching-fraction}, its fraction among diagrams is
therefore at most
\[
 \frac{|\mathcal X_\ell\cap\mathcal E_h|}{N_h}
 \leq(2h+1)\frac4{(4h)^2}\leq\frac4{4h}.
\]
On the other matchings $|f_\ell-\bar f_\ell|\leq t_\ell$, while
on the exceptional set it is at most two, since both
$|f_\ell|$ and $|\bar f_\ell|$ are at most one. Splitting the finite
sum into these two parts gives an average deviation at most
$t_\ell+2\cdot4/(4h)$. Together with
\eqref{eq:fixed-mode-full-mean}, this proves
\begin{equation}
 \frac1{N_h}\sum_{\rho\in\mathcal E_h}|f_\ell(\rho)|
 \leq\frac{6\ell^2}{4h}
 +16\ell\sqrt{\frac{\log(4h)}{4h}}+\frac8{4h},
 \qquad 1\leq\ell\leq h.
 \label{eq:fixed-mode-restricted-correlation}
\end{equation}
The restriction step uses only the cardinality of $\mathcal E_h$;
the single-cycle property was used earlier to identify diagrams and
to establish \eqref{eq:fixed-mode-orbit-reduction}.

\medskip
\noindent\emph{Decay at large genus and the genus sum.}
Substituting \eqref{eq:fixed-mode-restricted-correlation} into
\eqref{eq:fixed-mode-orbit-reduction}, and summing the elementary
polynomials in $\ell$, yields, for $1\leq K\leq h$,
\begin{equation}
 \frac1{N_h}\sum_{D\in\mathcal D_h}|B_D(m)|^2
 \leq\frac1K+\frac{K^2}{4h}
 +\frac{16}{3}K\sqrt{\frac{\log(4h)}{4h}}+\frac8{4h}.
 \label{eq:fixed-mode-grid-bound}
\end{equation}
Choose $K=\lfloor(4h/\log(4h))^{1/4}\rfloor$, which is a positive
integer at most $h$. Combining this estimate with
\eqref{eq:fixed-mode-length-error} and
$|I_{D,\bm x}(m,m)|^2\leq2|B_D(m)|^2+
2|I_{D,\bm x}(m,m)-B_D(m)|^2$ gives
\begin{equation}
 r_h(m)=O\!\left(
 \left(\frac{\log(4h)}{4h}\right)^{1/4}
 +\frac{m^2+1}{4h}\right),\qquad 4h>|m|.
 \label{eq:fixed-mode-genus-decay}
\end{equation}
The implied constant is absolute. In particular, $r_h(m)\to0$ as
$h\to\infty$ for every fixed nonzero mode $m$.

Finally, the weights in \eqref{eq:Z-definition} can be written as
\[
 w_h(g)=\frac{(g/2)^{2h+1}}{(2h+1)!\sinh(g/2)}.
\]
Their mass at $h<g/8$ is exponentially small. Indeed, for $g\geq8$
and $h<g/8$, one has $2h+1\leq3g/8$. For every integer
$0\leq j\leq3g/8$, multiply the corresponding exponential-series
term by the upper bound $1\leq e^{(3g/8-j)/4}$ and then sum. This gives
\begin{equation}
 \sum_{0\leq h<g/8}w_h(g)
 \leq\frac{\displaystyle\sum_{0\leq j\leq3g/8}(g/2)^j/j!}
 {\sinh(g/2)}
 \leq\frac{e^{3g/32+(g/2)e^{-1/4}}}{\sinh(g/2)}
 \leq3e^{-g/64},
 \label{eq:fixed-mode-genus-tail}
\end{equation}
where the last inequality uses $e^{-1/4}\leq25/32$ and
$\sinh(g/2)\geq e^{g/2}/3$ for $g\geq8$.
For $h\geq g/8$, one has $4h\geq g/2$, so for sufficiently large
$g$ the bound \eqref{eq:fixed-mode-genus-decay} applies uniformly
throughout this range. Using $0\leq r_h(m)\leq1$ in the remaining
range, the mixture \eqref{eq:rh-probability-mixture} satisfies
\begin{equation}
 p_{m,m}(g)
 \leq3e^{-g/64}
 +O\!\left(\left(\frac{\log g}{g}\right)^{1/4}
 +\frac{m^2+1}{g}\right)
 =O_m\!\left(\left(\frac{\log g}{g}\right)^{1/4}\right).
 \label{eq:fixed-mode-diagonal-decay}
\end{equation}
Equation \eqref{eq:entropy-diagonal-lower-bound} now proves
\eqref{eq:fixed-mode-entropy-bound}. The mode $m$ has remained fixed
throughout the proof.

\subsection{Monotonicity} \label{monotonicity}

We conjectured in \cite{Huang:2019uue} that for a fixed initial mode $m \neq0$,  the entropy is a monotonically  increasing function of the coupling $g>0$. 
Here we make a small progress by proving the monotonicity in a finite interval $0<g<  5.71$.  

Write the genus
contribution in the form
\begin{equation}
 A_h(m,n)=a^{(h)}_{m,n}g^{2h},\qquad
 \mathcal A_{m,n}(g):=\sum_{h\geq0}A_h(m,n)
 =\sum_{h\geq0}a^{(h)}_{m,n}g^{2h}.
 \label{eq:monotonicity-coefficient-expansion}
\end{equation}
For the two-mode correlators considered here, the genus-by-genus
non-negativity gives $a^{(h)}_{m,n}\geq0$, and
$a^{(0)}_{m,n}=\delta_{mn}$.  With $Z(g)$ from
\eqref{eq:Z-definition}, the probability and normalization are
\begin{equation}
 p_{m,n}(g)=\frac{\mathcal A_{m,n}(g)}{Z(g)},\qquad
 \sum_{n\in\mathbb Z}\mathcal A_{m,n}(g)=Z(g).
 \label{eq:monotonicity-normalization}
\end{equation}

For $n\neq m$, the planar coefficient vanishes.  Due to the uniform bound established in \cite{Huang:2019uue}, we can differentiate the entropy at each term in the infinite sum. Hence
\begin{equation}
 g\,\partial_g\mathcal A_{m,n}(g)
 =\sum_{h\geq1}2h\,a^{(h)}_{m,n}g^{2h}
 \geq2\mathcal A_{m,n}(g).
 \label{eq:monotonicity-offdiag-coefficient-bound}
\end{equation}
Since
\[
 \partial_g\log Z(g)
 =\frac12\coth\frac g2-\frac1g,
\]
we obtain
\begin{align}
 p_{m,n}'(g)
 &=\frac1{Z(g)}
 \left(\partial_g\mathcal A_{m,n}(g)
       -\mathcal A_{m,n}(g)\partial_g\log Z(g)\right) \notag\\
 &\geq \frac{p_{m,n}(g)}{g}
 \left(3-\frac g2\coth\frac g2\right),
 \qquad n\neq m .
 \label{eq:monotonicity-offdiag-derivative}
\end{align}
Let $g_{\rm off}$ be the positive solution of
\begin{equation}
 \frac{g_{\rm off}}2\coth\frac{g_{\rm off}}2=3,
 \qquad
 g_{\rm off}=5.969409\ldots .
 \label{eq:monotonicity-g-off}
\end{equation}
The function on the left is strictly increasing, so
$p_{m,n}'(g)>0$ for every nonzero off-diagonal probability whenever
$0<g<g_{\rm off}$.
It is easy to see from genus one formula that each $m\neq0$ has at least one nonzero off-diagonal transition probability, so the 
entropy monotonicity is strict rather than weak.

It remains to show that the diagonal probability is larger than every
off-diagonal probability.  The genus-one diagonal coefficient is
\begin{equation}
 a^{(1)}_{m,m}
 =\frac1{60}-\frac1{24\pi^2m^2}
  +\frac7{16\pi^4m^4},\qquad m\neq0 .
 \label{eq:monotonicity-genus-one-diagonal}
\end{equation}
For integer $m\neq0$ its minimum occurs at $|m|=2$:
\begin{equation}
 a^{(1)}_{m,m}\geq
 a_*:=\frac1{60}-\frac1{96\pi^2}
       +\frac7{256\pi^4}
 =0.0158919481\ldots .
 \label{eq:monotonicity-a-star}
\end{equation}
Non-negativity of all higher-genus diagonal coefficients therefore gives
\begin{equation}
 \mathcal A_{m,m}(g)\geq1+a_*g^2.
 \label{eq:monotonicity-diagonal-lower-bound}
\end{equation}
Let $g_*$ be the first positive solution of
\begin{equation}
 Z(g_*)=2\bigl(1+a_*g_*^2\bigr),
 \qquad
 g_*=5.7156157\ldots .
 \label{eq:monotonicity-g-star}
\end{equation}
This root is unique. Indeed,
\[
 \frac{Z(g)-2}{g^2}
 =-\frac1{g^2}+\frac1{24}
 +\sum_{h\geq2}\frac{g^{2h-2}}{2^{2h}(2h+1)!}
\]
is strictly increasing for $g>0$, with limits $-\infty$ at zero
and $+\infty$ at infinity. It therefore equals $2a_*$ exactly once,
and $Z(g)<2(1+a_*g^2)$ precisely for $0<g<g_*$. Together with
\eqref{eq:monotonicity-diagonal-lower-bound}, this gives
\begin{equation}
 2\mathcal A_{m,m}(g)>Z(g)
 =\mathcal A_{m,m}(g)+\sum_{n\neq m}\mathcal A_{m,n}(g),
 \label{eq:monotonicity-diagonal-dominance}
\end{equation}
and hence
$\mathcal A_{m,m}(g)>\mathcal A_{m,n}(g)$ for every $n\neq m$.
Equivalently, $p_{m,m}(g)>p_{m,n}(g)$.

 Using $\sum_np_{m,n}'(g)=0$, we can write
the entropy derivative as a sum over the nonzero off-diagonal probabilities
\begin{equation}
 S_m'(g)
 =\sum_{n\neq m}p_{m,n}'(g)
   \log\frac{p_{m,m}(g)}{p_{m,n}(g)} .
 \label{eq:monotonicity-entropy-derivative}
\end{equation}
For $0<g<g_*<g_{\rm off}$, every nonzero term
on the right-hand side of \eqref{eq:monotonicity-entropy-derivative} is
then strictly positive: the derivative is positive by
\eqref{eq:monotonicity-offdiag-derivative}, and the logarithm is positive by
\eqref{eq:monotonicity-diagonal-dominance}.  Thus
\begin{equation}
 S_m'(g)>0
 \qquad (m\neq0,\qquad 0<g<g_*=5.7156157\ldots),
 \label{eq:monotonicity-finite-range}
\end{equation}
so in particular the entropy is strictly increasing for
$0<g<5.71$.  

With AI assistance, one can further improve the result with more complicated arguments.  However for a fixed nonzero initial mode, no rigorous all-coupling proof or counterexample is known.

\section{Conclusion} 

We have made progress on several mathematically well-defined problems in the study of BMN strings. However, despite many prompts, the non-negativity conjecture (\ref{conjecture1}) in Section \ref{secnonnegative} and the all-coupling entropy monotonicity conjecture in Section \ref{monotonicity} remain unresolved. In particular, the non-negativity conjecture is a simple and well-tested prediction of holography, related to unitarity on the string theory side, so has a clear important physical meaning that warrants further investigation. Further analytical works, aided by better AI tools and more computational resources, may help resolve these questions.

\vspace{0.2in} {\leftline {\bf Acknowledgments}}
\nopagebreak
The derivations and proofs in this paper were developed with assistance from GPT-6 Astra. The text was written by the author with assistance from AI tools. The author is responsible for checking the arguments and for the final manuscript. This work is  supported by the National Natural Science Foundation of China Grants No. 12325502 and No. 12247103.  

\appendix

\section{A stronger conjecture}   \label{appendixcounter}

We consider an even stronger conjecture that, for every genus $h\geq1$
and every string diagram $i$ from the factorization formulas with two or three distinct string modes,
\begin{equation}
 S_i(\bm m,\bm n)\geq0 ? 
 \label{eq:string-diagram-positivity-conjecture}
\end{equation}
If this were true, then together with \eqref{eq:total-string-diagram-sum}, this would imply the non-negativity (\ref{conjecture1}) of the total two-point function separately at each genus.

\subsection{The tests} 
For the case of two string modes, we argued in \cite{Huang:2019lso} that most string diagram contributions with any intermediate states are non-negative, unless there is a sub-diagram containing a transition between a zero-mode string and a non-zero-mode string.  However, this is not complete as there are diagrams with an odd number of negative vertices that cannot be described in this way.  We can instead give a simple proof from the factorization formula. 
For fixed genus $h\geq1$, write $\bm m=(-m,m)$ and $\bm n=(-n,n)$.
For field diagram $j$, let $I_{j,r}$ be its $4h$ incoming intervals and
$P_{j,\bm x}$ the piecewise translation to the outgoing circle. Set
\[
 A_{j,r}:=\int_{I_{j,r}}
 e^{2\pi i[nP_{j,\bm x}(y)-my]}\,\dd y,
 \qquad A_j:=\sum_{r=1}^{4h}A_{j,r}.
\]
The two impurity phases are complex conjugates. With the first impurity
in the first interval, the field-diagram contribution is
$F_j=\frac{g^{2h}}{(4h-1)!}
\int_{\Delta_{4h-1}}\overline{A_{j,1}}A_j\,\dd\mu_h$,
using \eqref{eq:normalized-simplex-measure}.
Cyclically choosing another interval as the first preserves the simplex
measure and $m_{ij}$: relabeling the segments gives a bijection of the
splitting-and-joining histories. The common phase from shifting the circle
origins cancels between the two impurity factors. Averaging
\eqref{eq:general-factorization-proof} over these $4h$ choices therefore gives
\begin{equation}
\begin{aligned}
 S_i(\bm m,\bm n)
 &=\frac{g^{2h}}{(4h)!}\sum_jm_{ij}
 \int_{\Delta_{4h-1}}
 \left(\sum_{r=1}^{4h}\overline{A_{j,r}}\right)A_j\,\dd\mu_h\\
 &=\frac{g^{2h}}{(4h)!}\sum_jm_{ij}
 \int_{\Delta_{4h-1}}|A_j|^2\,\dd\mu_h\geq0.
\end{aligned}
\label{eq:two-mode-string-diagram-positivity}
\end{equation}
This proves non-negativity for each string diagram separately, with all
intermediate states and impurity routings summed, for the case of two string modes.

As discussed in \cite{Du:2021spv}, for two string modes each cyclic group
of field-theory diagrams is separately non-negative, as seen in \eqref{eq:two-mode-string-diagram-positivity}.
For three string modes,  a more detailed genus-two check finds negative
contributions in at least two of the four cyclic groups. At genus three
and higher, cyclic groups need not be closed under the two involutions,
so their contributions need not be real or symmetric under
${\bm m}\leftrightarrow{\bm n}$.  Since the cubic vertices are always real, a string diagram contribution is 
always real, but not necessarily symmetric, for any number of string modes.

Now we consider the non-trivial case of three string modes.  We will use exact computations to establish  that every individual string-diagram contribution at genus two  is non-negative for all mode numbers.  We then test individual non-reducible diagrams at genera three, four, and five.  The total computational time is less than an hour with AI tools. 

Here a string diagram specifies the directed connections of the cubic vertices,
and its contribution includes all allowed impurity routings and all
intermediate-state sums, as in Section~2.2.  We write
$S_i(\bm m,\bm n)$ when displaying its external-mode dependence.

\begin{figure}[htbp]
 \centering
 \includegraphics[width=\textwidth]{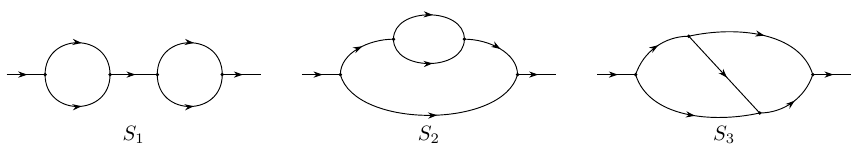}
 \caption{The three genus-two string diagrams.  The arrows run from
 the incoming string on the left to the outgoing string on the right;
 dots denote cubic vertices.  The diagram $S_1$ is reducible, whereas
 $S_2$ and $S_3$ are non-reducible.  Each contribution includes the sum
 over all allowed impurity routings and intermediate states.}
 \label{fig:genus-two-string-diagrams}
\end{figure}

The three genus-two string diagrams are shown in
Figure~\ref{fig:genus-two-string-diagrams}, with the same labeling as
in \cite{Huang:2010ne}.  Their vertices have unique compatible orders,
so \eqref{eq:total-string-diagram-sum} becomes
\begin{equation}
 \big\langle\overline{O}_{\bm m}O_{\bm n}\big\rangle_2
 =\frac{S_1(\bm m,\bm n)+S_2(\bm m,\bm n)
             +S_3(\bm m,\bm n)}{24}.
 \label{eq:genus-two-string-sum}
\end{equation}

We call a two-point string diagram \emph{reducible} if it can be
constructed by pasting the outgoing external string of one
positive-genus two-point diagram to the incoming external string of
another, without cutting any internal string of either diagram.
Otherwise it is \emph{non-reducible}.  Equivalently, a reducible
diagram has an intermediate single string that separates all earlier
vertices from all later vertices.  In Figure~\ref{fig:genus-two-string-diagrams},
$S_1$ is the paste of two genus-one diagrams.  Although $S_2$ contains
a genus-one insertion on one branch, producing it in this way requires
cutting an internal string of the outer diagram; hence $S_2$ is
non-reducible under this definition.

Pasting gives the complete sum over intermediate single-string states.
For example, if $S(\bm m,\bm n)$ denotes the genus-one string
contribution of Section~2.1, then
\begin{equation}
 S_1(\bm m,\bm n)
 =\sum_{\substack{\bm p\in\Z^3\\p_1+p_2+p_3=0}}
 S(\bm m,\bm p)S(\bm p,\bm n)\geq0,
 \label{eq:reducible-string-positivity}
\end{equation}
by the established genus-one non-negativity.  The same argument applies
inductively to every reducible diagram once the lower-genus
contributions are non-negative.  It is therefore enough to establish
non-negativity for the non-reducible diagrams.

\medskip
\noindent\emph{A computer-assisted proof at genus two.}
The two non-reducible genus-two contributions satisfy
$S_2(\bm m,\bm n)\geq0$ and $S_3(\bm m,\bm n)\geq0$ for every
level-matched integer pair.  We prove this by evaluating the finite
factorization sums \eqref{eq:general-factorization-proof} exactly and
certifying the signs of the resulting rational expressions.  Away from
$m_a=0$, $n_a=0$, and $m_a\pm n_b=0$, the result takes the form
\begin{equation}
 S_i=g^4\left(\frac{C_{i,4}}{X^2}
             -\frac{C_{i,6}}{X^3}
             +\frac{C_{i,8}}{X^4}\right),
 \qquad X=(2\pi)^2,\qquad i=2,3,
 \label{eq:genus-two-exact-generic}
\end{equation}
where the $C_{i,k}$ are rational functions of the external modes.
Writing $d_a=m_a-n_a$, the leading coefficients are particularly simple:
\begin{equation}
 C_{2,4}=\frac19\sum_{a<b}\frac{1}{d_a^2d_b^2}>0,
 \qquad
 C_{3,4}=\frac15\sum_{a<b}\frac{1}{d_a^2d_b^2}>0.
 \label{eq:genus-two-leading-coefficients}
\end{equation}

By a brute force analysis of the complicated rational functions, AI tools can exhaust all possible cases and prove the two string diagram contributions in  (\eqref{eq:genus-two-exact-generic}) are non-negative for all mode numbers. However this is not very illuminating for understanding the conjecture at a general higher genus and we skip the detailed arguments here.

\medskip
\noindent\emph{Tests at higher genus.}
We evaluate the individual contributions in
\eqref{eq:general-factorization-proof} for all $17$ non-reducible
genus-three diagrams and all $203$ non-reducible genus-four diagrams.
At genus five we test six selected diagrams among the $3{,}520$
non-reducible topologies ($4{,}066$ topologies in total).  These six
are obtained by inserting selected genus-four two-point diagrams into
one branch of a genus-one diagram.  The untouched branch rules out an
intermediate single-string separator, so these diagrams are
non-reducible under the definition above.  Deleting the outer splitting
and joining vertices reduces their multiplicity calculation to genus
four.  We checked this reduction against complete direct enumeration
through genus four.  Each selected genus-five contribution
includes the complete sum over field diagrams with the multiplicities
$m_{ij}$ and all impurity routings. The tests are summarized in 
Table~\ref{tab:string-diagram-positivity-tests}. We note that these are 
approximate numerical tests, which are not  exact tests but can be performed much faster.

\begin{table}[htbp]
 \centering
 \begin{tabular}{c r r r r}
  \hline
  Genus & Non-reducible & Mode pairs & Separate & Negative \\
  $h$ & diagrams tested & per diagram & evaluations & results \\
  \hline
  $3$ & all $17$  & $144{,}960$ & $2{,}464{,}320$ & $0$ \\
  $4$ & all $203$ & $3{,}032$ & $615{,}496$ & $0$ \\
  $5$ & $6$ of $3{,}520$ & $1{,}507$ & $9{,}042$ & $0$ \\
  \hline
 \end{tabular}
 \caption{Higher-genus tests of individual non-reducible string-diagram
 contributions.  Every indicated diagram is evaluated separately at
 each listed mode pair, using its complete factorization sum.}
 \label{tab:string-diagram-positivity-tests}
\end{table}

\subsection{A counterexample}

Surprisingly, although the conjecture \eqref{eq:string-diagram-positivity-conjecture} passes extensive tests at low genera, with AI assistance, one can construct an ingenious  counterexample at asymptotically large genus. 

\begin{figure}[htbp]
 \centering
 \includegraphics[width=\textwidth]{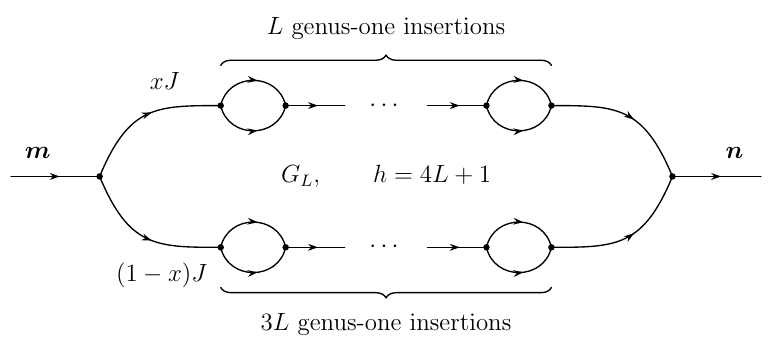}
 \caption{The counterexample family $G_L$, of genus $h=4L+1$.
 The two daughters have sizes $xJ$ and $(1-x)J$ and carry respectively
 $L$ and $3L$ consecutive genus-one insertions, indicated schematically
 by the ellipses. Arrows run from the incoming to the outgoing string;
 dots denote cubic vertices. The contribution includes integration over
 $x$ and sums over all intermediate modes and impurity routings.}
 \label{fig:counterexample-diagram}
\end{figure}

For a positive integer $L$, let $G_L$ be the diagram shown in
Figure~\ref{fig:counterexample-diagram}, obtained by splitting the incoming
string into two daughters, inserting $L$
consecutive genus-one two-point diagrams on the first daughter and
$3L$ on the second, and finally joining the daughters. Its genus is
$h=4L+1$. Both branches persist between the outer splitting and joining
vertices, so $G_L$ is non-reducible in the sense defined above. We will
show that, with all intermediate states and impurity routings included,
\begin{equation}
 \lim_{L\to\infty}
 \frac{S_{G_L}((1,2,-3),(2,3,-5))}
      {S_{G_L}((0,0,0),(0,0,0))}
 =-\frac{256}{1215\pi^6}<0.
 \label{eq:counterexample-limit}
\end{equation}
The denominator is strictly positive for $g>0$. Thus the numerator is
negative at every sufficiently large finite $L$, although the argument
does not determine the first negative genus.

\medskip
\noindent\emph{Repeated genus-one insertions.}
For $0\leq r\leq3$, let $S_r^{(1)}$ denote the genus-one string
contribution \eqref{tag3} with $r$ labeled impurities, and define
\begin{equation}
 T_r(\bm m,\bm n)
 :=\frac{12}{g^2}S_r^{(1)}(\bm m,\bm n)
 =\frac{24}{g^2}
 \big\langle\overline O_{\bm m}O_{\bm n}\big\rangle_1.
 \label{eq:counterexample-kernel}
\end{equation}
The operator acts on the Hilbert space
$\mathcal H_r=\ell^2(\Lambda_r)$, where
$\Lambda_r=\{\bm m\in\Z^r:\sum_a m_a=0\}$:
\begin{equation}
 (T_rf)(\bm m)=\sum_{\bm n\in\Lambda_r}
 T_r(\bm m,\bm n)f(\bm n),\qquad
 \langle f,u\rangle=\sum_{\bm m\in\Lambda_r}
 \overline{f(\bm m)}u(\bm m).
 \label{eq:counterexample-Hilbert-space}
\end{equation}
Here $f$ is a square-summable sequence of mode coefficients. The tuples
are ordered by the impurity labels, with zero and repeated integer
entries allowed; there is no quotient by permutations. For $r=0$ the
only tuple is the empty tuple, and for $r=1$ it is $(0)$, so these two
spaces are one-dimensional.

The established genus-one non-negativity and normalization
\cite{Du:2021spv} give
\[
 T_r(\bm m,\bm n)=T_r(\bm n,\bm m)\geq0,
 \qquad \sum_{\bm n}T_r(\bm m,\bm n)=1.
\]
Symmetry gives unit column sums as well. Weighted Cauchy--Schwarz then
implies $\|T_rf\|^2\leq\sum_{\bm m,\bm n}
T_r(\bm m,\bm n)|f(\bm n)|^2=\|f\|^2$.
Moreover, the split--join formula \eqref{tag3} gives, initially for
finitely supported $f$,
\begin{equation}
 \langle f,T_rf\rangle
 =\frac{6J}{g^2}\int_0^1\dd x
 \sum_{A,\bm p,\bm q}
 \left|\sum_{\bm m}f(\bm m)
 \overline{V_{\bm m}^{A;\bm p,\bm q}(x)}\right|^2\geq0,
 \label{eq:counterexample-Gram}
\end{equation}
with the routing and level-matching sums of \eqref{tag3}.
By continuity this holds for all $f\in\mathcal H_r$. Hence
$0\leq T_r\leq I$ as operators. This operator positivity is a separate
property from the non-negativity of the individual matrix elements.

We next determine the fixed vectors. The all-zero state
$e_{\bm0}(\bm m)=\delta_{\bm m,\bm0}$ is decoupled:
$T_r(\bm m,\bm0)=\delta_{\bm m,\bm0}$, as follows directly from
\eqref{tag8}. If $T_rf=f$, symmetry and the unit sums give
\begin{equation}
 0=\langle f,(I-T_r)f\rangle
 =\frac12\sum_{\bm m,\bm n}
 T_r(\bm m,\bm n)|f(\bm m)-f(\bm n)|^2.
 \label{eq:counterexample-fixed-vectors}
\end{equation}
Every nonzero tuple $\bm m$ has infinitely many $\bm n$ for which
$T_r(\bm m,\bm n)>0$. For three nonzero entries, take
$\bm n=(N,2N,-3N)$ with $N>\max_a|m_a|$; the generic genus-one
formula \cite{Du:2021spv} gives
\[
 T_3(\bm m,\bm n)
 =\frac{3\sum_a(m_a-n_a)^2}
 {4\pi^4\prod_a(m_a-n_a)^2}>0.
\]
If a labeled mode vanishes on both sides, its factor in
\eqref{tag8} is one, so the kernel reduces to the two-mode kernel.
For a fixed $m\neq0$, the two-mode formula \cite{Huang:2019uue} gives
\[
 \lim_{n\to\infty}n^2
 T_2((-m,m),(-n,n))
 =\frac{2}{\pi^2}+\frac{6}{\pi^4m^2}>0.
\]
Equation~\eqref{eq:counterexample-fixed-vectors} therefore forces
$f$ to take the same value at infinitely many tuples whenever
$f(\bm m)\neq0$ for a nonzero $\bm m$. Square summability excludes
this possibility. Thus $\ker(I-T_r)=\mathbb C e_{\bm0}$.

The existence of the limit of $T_r^Lf$ also follows from
$0\leq T_r\leq I$, without assuming a spectral gap. Indeed,
$0\leq T_r^{L+1}\leq T_r^L\leq I$, so
$a_L:=\langle f,T_r^Lf\rangle$ decreases to a finite limit.
For $M>L$, the operator $D=T_r^L-T_r^M$ satisfies $0\leq D\leq I$,
and therefore
\[
 \|T_r^Lf-T_r^Mf\|^2
 =\langle f,D^2f\rangle
 \leq\langle f,Df\rangle=a_L-a_M\longrightarrow0.
\]
Thus the sequence is Cauchy and converges in $\mathcal H_r$.
The limit is fixed by $T_r$ and preserves the inner product with
every fixed vector, so it is the orthogonal projection onto
$\ker(I-T_r)$. In particular,
\begin{equation}
 T_r^L\longrightarrow P_{0,r}\quad\hbox{strongly},\qquad
 P_{0,r}f=f(\bm0)e_{\bm0},\qquad
 \lim_{L\to\infty}\|T_r^Lf-f(\bm0)e_{\bm0}\|_2=0.
 \label{eq:counterexample-strong-limit}
\end{equation}
Strong convergence means convergence for each fixed square-summable
$f$, with no uniform rate over all unit vectors asserted. The
all-zero vector is a delta sequence, not the constant sequence over
mode labels. For a nonzero initial mode, the ordinary sum of the
matrix elements of $T_r^L$ remains one while their squared sum tends
to zero: the distribution spreads through infinitely many nonzero
modes, rather than flowing into the decoupled zero mode.

\medskip
\noindent\emph{The complete diagram and its large-$L$ limit.}
Let the two outer daughters have sizes $xJ$ and $(1-x)J$, and retain
$A,B,a,b$ from Section~2.1, now with $a+b=3$. Remove the common
coupling factor from the cubic vertex by defining the vector
\[
 v_{\bm m,A}(x)_{\bm p,\bm q}
 :=\frac{\sqrt J}{g}V_{\bm m}^{A;\bm p,\bm q}(x)
 \quad\hbox{in }\mathcal H_a\otimes\mathcal H_b.
\]
The effective coupling on the first daughter is $gx^2$, so each
genus-one insertion there contributes $(g^2x^4/12)T_a$; the second
daughter similarly contributes $(g^2(1-x)^4/12)T_b$. Consequently
\begin{align}
 S_{G_L}(\bm m,\bm n)
 &=c_L(g)\int_0^1 x^{4L}(1-x)^{12L}\Phi_L(x)\,\dd x,
 \nonumber\\
 \Phi_L(x)
 &:=\sum_{A\subseteq\{1,2,3\}}
 \left\langle v_{\bm n,A}(x),
 (T_a^L\otimes T_b^{3L})v_{\bm m,A}(x)\right\rangle,
 \label{eq:counterexample-sewing}
\end{align}
where $c_L(g)>0$ is independent of the external modes and includes
the common coupling and symmetry factors. All intermediate-state
sums are contained in the operator powers and the inner products.

For all-zero external modes, the vertex vector has only its
all-zero daughter component, equal to
$x^{(a+1)/2}(1-x)^{(b+1)/2}$. Thus
$\Phi_L(x)=\sum_A x^{a+1}(1-x)^{b+1}=x(1-x)$ in this case.
For general external modes, Parseval applied to \eqref{tag2}, first
summing over unrestricted daughter Fourier modes, gives
\[
 \|v_{\bm m,A}(x)\|^2
 \leq x^{a+1}(1-x)^{b+1}.
\]
Since the operator powers are contractions, it follows that
$|\Phi_L(x)|\leq x(1-x)$. Setting
$H_L(x)=\Phi_L(x)/[x(1-x)]$, we obtain the exact ratio
\begin{equation}
 \frac{S_{G_L}(\bm m,\bm n)}{S_{G_L}(\bm0,\bm0)}
 =\frac{\displaystyle\int_0^1
 x^{4L+1}(1-x)^{12L+1}H_L(x)\,\dd x}
 {\mathrm B(4L+2,12L+2)},\qquad |H_L(x)|\leq1,
 \label{eq:counterexample-beta-average}
\end{equation}
where $\mathrm B(s,t)=\int_0^1x^{s-1}(1-x)^{t-1}\dd x$.
In particular, the denominator of the ratio is strictly positive.

For external triples with no zero entries, put
\[
 C_{\bm m}(x):=\prod_{j=1}^3
 \frac{\sin(\pi m_jx)}{\pi m_j}.
\]
The all-zero daughter component of \eqref{tag2} is
\[
 v_{\bm m,A}(x)_{\bm0,\bm0}
 =(-1)^b x^{(1-a)/2}(1-x)^{(1-b)/2}C_{\bm m}(x).
\]
Here each integral over $[x,1]$ is minus the integral over $[0,x]$,
and level matching cancels the common phase. Using
\eqref{eq:counterexample-strong-limit} and summing over every routing
therefore gives
\begin{equation}
 H_L(x)\longrightarrow H_\infty(x)
 :=\frac{C_{\bm m}(x)C_{\bm n}(x)}{x^3(1-x)^3}.
 \label{eq:counterexample-zero-mode-limit}
\end{equation}
Indeed, the routing sum of the products of the zero-mode components
is $C_{\bm m}(x)C_{\bm n}(x)/[x^2(1-x)^2]$.
This convergence is uniform on every closed subinterval of $(0,1)$:
after rescaling the daughter intervals to unit length, the Fourier
coefficient vectors $v_{\bm m,A}(x)$ are continuous in Hilbert-space
norm, and strongly convergent contractions converge uniformly on
compact sets of vectors.

The normalized weight in \eqref{eq:counterexample-beta-average}
concentrates at $x=1/4$, the unique maximum of $x(1-x)^3$.
The local uniform convergence just established and the bound
$|H_L|\leq1$ thus imply that the ratio tends to $H_\infty(1/4)$.
Finally,
\[
 C_{(1,2,-3)}(1/4)=\frac{1}{12\pi^3},\qquad
 C_{(2,3,-5)}(1/4)=-\frac{1}{60\pi^3},
\]
which proves \eqref{eq:counterexample-limit}. Repeated insertions
select the daughter zero modes in the above Hilbert-space sense,
while the unequal numbers of insertions select a size ratio where
the two external cubic couplings have opposite signs. This rules
out \eqref{eq:string-diagram-positivity-conjecture}, but does not
contradict the fixed-genus positivity conjecture \eqref{conjecture1},
which concerns the complete weighted sum
$\sum_i L_iS_i/(2h)!$ over all string diagrams.

\clearpage

\section{Genus three string diagrams and their weights}  \label{genusthree}

\begin{figure}[p]
 \centering
 \includegraphics[width=\textwidth]{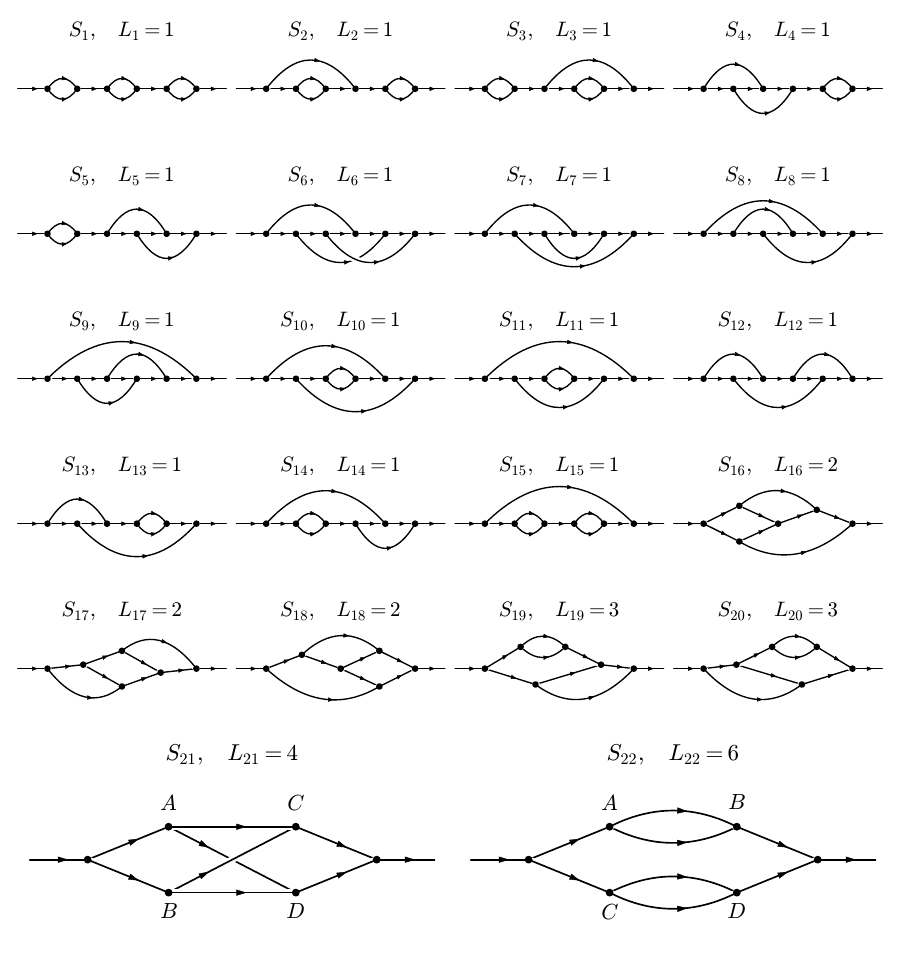}
 \caption{All 22 genus-three string diagrams and their weights $L_i$.
 Arrows run from the incoming string on the left to the outgoing
 string on the right.  Black dots denote cubic vertices; crossings
 without dots do not join the strings.  Parallel curves represent
 separate strings.  The four middle vertices of $S_{21}$ and
 $S_{22}$ are marked $A,B,C,D$ for the counting in the text.
 The initial split and final join are unmarked because their
 positions in the ordering are fixed.}
 \label{fig:genus-three-string-diagrams}
\end{figure}

The 22 genus-three string diagrams are shown in
Figure~\ref{fig:genus-three-string-diagrams}, with their weights $L_i$
indicated in each panel.  The numbering $S_1,\ldots,S_{22}$ is local
to this appendix.  The diagrams $S_1,\ldots,S_5$ are reducible, while
the other 17 are non-reducible in the sense of
Appendix~\ref{appendixcounter}.  With all intermediate states and
impurity routings included in each $S_i$, the genus-three case of
\eqref{eq:total-string-diagram-sum} is
\begin{equation}
 \begin{aligned}
 \big\langle\overline{O}_{\bm m}O_{\bm n}\big\rangle_3
 =\frac{1}{720}\Big[&\sum_{i=1}^{15}S_i
       +2(S_{16}+S_{17}+S_{18})\\
       &+3(S_{19}+S_{20})+4S_{21}+6S_{22}\Big].
 \end{aligned}
 \label{eq:genus-three-string-sum}
\end{equation}

To read $L_i$ from a graph, order its six cubic vertices so that
every vertex occurs after all vertices connected to it by an
incoming directed path.  The initial split is necessarily first
and the final join last, so only the order of the four middle
vertices remains to be counted.  The horizontal positions of
vertices on independent branches impose no additional ordering.
For $S_1,\ldots,S_{15}$, a directed path passes through all six
vertices, fixing their order and giving $L_i=1$.

The four middle vertices of $S_{21}$ and $S_{22}$ are marked
$A,B,C,D$ directly in Figure~\ref{fig:genus-three-string-diagrams}.
Writing $A<B$ when $A$ must occur before $B$, consider first
$S_{21}$.  Each joining vertex $C,D$ receives a string from each
splitting vertex $A,B$, so
\begin{equation}
 A<C,\qquad A<D,\qquad B<C,\qquad B<D.
 \label{eq:genus-three-S21-order}
\end{equation}
The splits $A,B$ can occur in either order, followed by the joins
$C,D$ in either order.  Omitting the initial and final cubic
vertices, the four allowed orders are
\[
 ABCD,\qquad ABDC,\qquad BACD,\qquad BADC.
\]
Therefore $L_{21}=2!\,2!=4$.

For $S_{22}$, the upper bubble splits at $A$ and joins at $B$,
while the lower bubble splits at $C$ and joins at $D$.  The only
constraints on these four vertices are
\begin{equation}
 A<B,\qquad C<D.
 \label{eq:genus-three-S22-order}
\end{equation}
The two ordered pairs can be interleaved in six ways:
\[
 ABCD,\quad ACBD,\quad ACDB,\quad
 CABD,\quad CADB,\quad CDAB.
\]
Equivalently, choose which two of the four positions are occupied
by $A,B$; the internal orders of both pairs are then fixed.  Hence
$L_{22}=\binom{4}{2}=6$.  These orderings treat the vertices as
distinct events; exchanging the two branches does not divide
the count $L_i$.

\clearpage

\section{Unistochastic matrices and polygon closure}
\label{appendixunistochastic}

Let $a,b$ denote two distinct BMN mode tuples in
$\Lambda_k=\{\bm m\in\Z^k:\sum_i m_i=0\}$, with $k=2,3$.
If \eqref{probability} holds, write
$U_{aj}=\sqrt{p_{aj}}e^{i\theta_{aj}}$ and define the polygon side lengths
$s_j=\sqrt{p_{aj}p_{bj}}$. Orthogonality of the two rows requires
\[
 \sum_{j\in\Lambda_k}s_j e^{i(\theta_{aj}-\theta_{bj})}=0.
\]
This series is absolutely convergent, since Cauchy--Schwarz and row
normalization give $\sum_j s_j\leq1$. Isolating any one term and using
the triangle inequality gives $s_j\leq\sum_{\ell\ne j}s_\ell$ for
every $j$, or equivalently
\begin{equation}
 \boxed{2\sup_j\sqrt{p_{aj}p_{bj}}
 \leq\sum_j\sqrt{p_{aj}p_{bj}}.}
 \label{eq:unistochastic-polygon}
\end{equation}
Geometrically, no side of a closed polygon can be longer than the sum
of the remaining sides. These are also called the \emph{bracelet
conditions}; analogous conditions hold for every pair of columns.
For our symmetric matrix the row
and column conditions coincide. They are necessary, but do not in
general guarantee compatible phases for all row pairs. As a simple
example of a doubly stochastic matrix that is not unistochastic,
consider
\[
 \frac12\begin{pmatrix}1&1&0\\1&0&1\\0&1&1\end{pmatrix}
\]
whose first two rows have
only one nonzero overlap, so they violate
\eqref{eq:unistochastic-polygon}.

We now prove \eqref{eq:unistochastic-polygon} for every fixed pair
$a\ne b$ at sufficiently small $g$, by an argument applying to both
$k=2$ and $k=3$. For $k=3$ we assume the genus-by-genus non-negativity
conjecture \eqref{conjecture1}. The allowed interval of $g$ may depend
on $a,b$.
The set of labeled components that vanish is conserved by the
two-point function: for a measure-preserving interval exchange $T$,
\[
 I_T(m,n):=\int_0^1 e^{2\pi i[nT(y)-my]}\dd y,
 \qquad I_T(m,0)=I_T(0,m)=\delta_{m0}.
\]
Thus different zero-pattern sectors have disjoint row supports and
satisfy \eqref{eq:unistochastic-polygon} trivially at every coupling.
A common zero component contributes a factor one and reduces the
three-mode kernel to the two-mode kernel.

Within any nontrivial sector, the convergent genus expansion and the
explicit genus-one formulae, including their degenerate cases
\cite{Huang:2010ne,Huang:2019uue,Du:2021spv}, give
\[
 p_{ab}(g)=g^2q_{ab}+O(g^4),\qquad q_{ab}>0,
 \qquad p_{aa}(g)=1+O(g^2).
\]
Here $q_{ab}$ is the coefficient of $g^2$ in the genus-one correlator.
Choose any fixed third state $c\ne a,b$ in the same sector. The strict
positivity above ensures $q_{ac}q_{bc}>0$.
Symmetry $p_{ab}=p_{ba}$ then implies
\begin{align*}
 s_a+s_b&=2g\sqrt{q_{ab}}+O(g^3),\\
 |s_a-s_b|&=\sqrt{p_{ab}}\,
       |\sqrt{p_{aa}}-\sqrt{p_{bb}}|=O(g^3),\\
 s_c&=g^2\sqrt{q_{ac}q_{bc}}+O(g^4).
\end{align*}
Meanwhile, non-negativity and row normalization give
$p_{aj}\leq1-p_{aa}$ for $j\ne a$. The planar diagonal contribution
and genus-by-genus non-negativity also give $p_{aa}\geq1/Z(g)$, where
$Z(g)=2\sinh(g/2)/g$. Consequently,
\begin{equation}
 s_j\leq\sqrt{(1-p_{aa})(1-p_{bb})}
 \leq1-\frac1{Z(g)}=\frac{g^2}{24}+O(g^4),
 \qquad j\ne a,b.
 \label{eq:unistochastic-uniform-overlap}
\end{equation}
This bound is uniform over all intermediate columns. For sufficiently
small positive $g$, we have $|s_a-s_b|\leq s_c\leq\sum_{j\ne a,b}s_j$,
which proves the
polygon inequalities for the two endpoint columns $a,b$.
Equation~\eqref{eq:unistochastic-uniform-overlap} also gives
$s_j\leq s_a+s_b$ simultaneously for every other column, proving all
remaining inequalities. At $g=0$ they hold trivially since $P(0)=I$.
This establishes the necessary condition \eqref{eq:unistochastic-polygon} for each fixed pair for small coupling $g$. 

At stronger coupling, the transition probabilities are expected to
spread over an increasing number of final states. For two fixed
initial states in the same nontrivial conserved sector, suppose
their distributions broadly overlap over $N_{\mathrm{eff}}(g)\gg1$
states, with comparable probabilities of order
$1/N_{\mathrm{eff}}(g)$ and no larger peaks elsewhere. Then the
largest polygon side $s_j=\sqrt{p_{aj}p_{bj}}$ is of order
$1/N_{\mathrm{eff}}(g)$, while the total length $\sum_j s_j$ remains
of order one. Consequently,  the polygon condition
\eqref{eq:unistochastic-polygon} is satisfied comfortably with a large margin at
strong coupling. This is a heuristic based on the spreading of the
common overlap; increasing entropy alone does not establish the
condition. Some further partial results can be obtained, but a complete rigorous proof of this seeming quite weak 
condition \eqref{eq:unistochastic-polygon}  is surprisingly not so easy for AI models.

\phantomsection
\addcontentsline{toc}{section}{References}

\bibliographystyle{utphys} 
\bibliography{ReferenceBMN}


\end{document}